\documentclass[]{aa}
\usepackage{txfonts}
\usepackage{natbib}
\usepackage{graphicx}
\usepackage{aalongtable}
\usepackage{longtable}
\bibpunct{(}{)}{;}{a}{}{,}
\begin{document}

\title{A lack of close binaries among hot horizontal branch stars in globular
clusters \thanks{Based on observations with the ESO Very Large Telescope
at Paranal Observatory, Chile (proposal ID 078.D-0825)}
}
\subtitle{II. NGC\,2808}

\author{
C. Moni Bidin \inst{1}
\and
S. Villanova \inst{1}
\and
G. Piotto \inst{2}
\and
Y. Momany \inst{3,4}
}

\institute{
Departamento de Astronom\'ia, Universidad de Concepci\'on, Casilla 160-C, Concepci\'on, Chile
\and
Dipartimento di Astronomia, Universit\`{a} di Padova,
Vicolo dell'Osservatorio 3, 35122 Padova, Italy
\and
European Southern Observatory, Alonso de Cordova 3107, Vitacura, Santiago, Chile
\and
INAF- Oss. Astronomico di Padova, Vicolo dell'Osservatorio 5, 35122 Padova, Italy
}
\date{Received / Accepted }


\abstract
{The formation mechanism of hot horizontal branch (HB) stars is still one of the most uncertain points
of stellar evolution theories. In the past decade, models based on their binary origin have been very
successful in reproducing the properties of field subdwarf-B stars, but the observations of their
analogues in globular clusters has posed new problems. In addition, the discovery of multiple populations
offered an appealing alternative scenario for the formation of these stars.}
{We search for binaries of period $\wp\leq$200~days among a sample of 83 blue horizontal branch stars
(T$_\mathrm{eff}$=12\,000-22\,000~K) in NGC\,2808, a cluster known to host three distinct stellar
populations and a multimodal horizontal branch. The final sample, after the rejection of stars with
incomplete observations or poor quality data, consists of 64 targets.}
{The radial velocity of the targets was measured in fourteen epochs, spanning a temporal interval of
$\sim$75~days. The significant variations were identified by means of a detailed error analysis and a
statistical study.}
{We detect no RV variable object among stars cooler than the photometric G1 gap at $\sim$17\,000~K, while
two close ($\wp\leq$10 days) and two intermediate-period ($\wp$=10-50 days) systems are found among hotter
targets. The close and intermediate-period binary fraction for stars cooler than the gap are
$f_\mathrm{c}\leq$5\% and $f_\mathrm{ip}\leq$10\%, respectively, with 95\% confidence. The most probable
values among hotter stars are $f_\mathrm{c}\sim$20\% and $f_\mathrm{ip}\sim$30\%, but the
90\%- confidence level intervals are still large (6-42\% and 11-72\%, respectively).}
{The G1 gap appears as a discontinuity in the binary faction along the HB, with a higher incidence of
binaries among hotter stars, but a constant increase in $f$ with temperature rather than a discontinuity
cannot be excluded from our observations. We also find that intermediate-period binaries, never investigated
before among cluster HB stars, could play an important role among hotter stars, being more than $\sim$15-20\%
of the hottest stars of our sample. Our results, compared with previous estimates for other clusters, indicate
that $f_\mathrm{c}$ among hot HB stars is most probably higher for younger clusters, confirming the recently
proposed age-$f_\mathrm{c}$ relation.
However, the large observed difference in binary fraction between clusters (e.g. NGC\,2808 and NGC\,6752) is
still not reproduced by binary population synthesis models.
}


\authorrunning{Moni Bidin et al.}
\titlerunning{EHB binaries in NGC\,2808}
\maketitle


\section{Introduction}
\label{c_intro}

Horizontal branch (HB) stars in Galactic globular clusters (GCs) are old post- He flash stars of low initial
mass (0.7-0.9 M$_\odot$) that, after the exhaustion of hydrogen in the stellar core and their ascension along
the red giant branch, eventually ignited helium \citep{Hoyle55,Faulkner66}.

GCs display large differences in the HB morphology \citep[see, for example,][]{Piotto02}. The first parameter
responsible for this phenomenon is metallicity, but it alone cannot account for the complex observational picture
\citep[the so-called ''second parameter problem",][]{Sandage67,VanDenBergh67}. While some clusters contain
only red HB stars cooler than the RR-Lyrae gap, others host a large population of blue He-burning stars
extending even beyond the canonical end of the HB at $\sim$35\,000~K \citep[e.g.][]{Moehler04}. Even more puzzling,
in the color-magnitude diagram (CMD) of some clusters the HB appears continuous in its whole extension
\citep[e.g. NGC\,6752,][]{Momany02}, while in others it is clearly multimodal \citep[e.g. NGC\,2808,][]{Sosin97}.
This observational picture still lacks full comprehension, as a consequence of our poor understanding of the
formation mechanism of HB stars in GCs. The HB morphology has been linked, among others, to cluster age
\citep{Dotter10}, cluster concentration \citep{FusiPecci93}, stellar rotation \citep{Peterson83}, cluster
mass \citep{RecioBlanco06}, helium, and the environment of formation \citep{FraixBurnet09}, but none of the proposed
second parameters could satisfactorily reproduce the
complex observed behavior \citep[see][for a review]{Catelan09}. In particular, the most challenging task
is to account for the formation of extreme horizontal branch (EHB) stars at the faint hotter end of
HBs (T$_\mathrm{eff}\geq$20\,000~K), observed even in high metallicity clusters \citep[e.g.][]{Rich97} and
old open clusters \citep[NGC\,6791,][]{Buson06}. Stars
hotter than this critical temperature do not have an external envelope massive enough to sustain the shell
H-burning, and after the exhaustion of helium in the core they evolve directly to the white dwarf (WD) cooling
sequence, without ascending the asymptotic giant branch \citep[AGB manqu\'e stars,][]{Greggio90}. They are
extensively observed and studied in the Galactic field, identified as the so-called subdwarf B-type (sdB)
stars \citep{Greenstein71,Caloi72,Heber86}. However, in GCs they are still poorly studied because of their
faintness, and many questions still await an answer \citep[see][for recent reviews]{Catelan09,Moni10}. Many
single-star evolutionary channels have been invoked to explain EHB star formation in GCs, including interactions
with a close planet \citep[][see also \citealt{Silvotti07}]{Soker98}, helium mixing driven by either internal
rotation \citep{Sweigart79,Sweigart97} or stellar encounters \citep{Suda07}, and close encounters with a central
intermediate-mass black hole \citep{Miocchi07}.

The dynamical interactions inside binary systems were proposed early on to be responsible for the heavy mass-loss
required to form a EHB star \citep{Mengel76,Tutukov87}. In the past decade, the ''binary scenario" has achieved
many observational and theoretical successes among field sdB stars, and is now widely accepted as the most
satisfactory explanation of their formation. The binary population synthesis model of
\citet{Han02,Han03,Han07} could reproduce their observational properties in great detail, although a small
fraction of progenies of single stars is probably required for a perfect match \citep{Lisker05}. Han's model
considers three main formation channels: the stable Roche Lobe Overflow (RLOF), which produces sdB's in wide
binaries; the common envelope (CE) channel, which forms close systems; and the merging of two WDs, whose progenies
are single stars. On the other hand, many surveys have confirmed that a large fraction of field sdB
stars reside in binaries \citep{Ferguson84,Allard94,Ulla98,Aznar01,Maxted01,Williams01,Reed04,Napiwotzki04},
and sdB's in close systems with periods shorter than ten days are very common
\citep{Moran99,Saffer98,Heber02,MoralesRueda03}, although the exact close-binary fraction is still uncertain,
ranging from 40-45\% \citep{Napiwotzki04} to 70\% \citep{Maxted01}.

Observations of EHB stars in GCs have so far presented a challenge to the binary scenario, at variance with
its well-established successes for field stars. The first surveys surprisingly revealed a lack of EHB
close systems in GCs \citep{Moni06a,Moni06b}. \citet{Moni08a} fixed at 4\% the most probable value of the EHB
close-binary fraction ($f_\mathrm{c}$) in \object{NGC\,6752}, proposing that a decrease in $f_\mathrm{c}$ with the
age of the stellar population should be a natural expectation of the binary scenario. The detailed calculations of
\citet{Han08} confirmed that the CE channel becomes very inefficient after the first few Gyrs, and the WD-WD
merging should be the predominant mechanism for the formation of EHB stars in old stellar systems. As a
consequence, cluster EHBs should be principally single-star products of merging, and close systems should be
rare. However, \citet{Moni09} measured a higher $f_\mathrm{c}$ in \object{M\,80} and \object{NGC\,5986}
(12\% and 25\%, respectively), and argued that the models proposed by \citet{Han08} cannot simultaneously account
for the low $f_\mathrm{c}$ measured in \object{NGC\,6752} and these much higher values found in clusters only
1-2 Gyr younger. Unfortunately, their results suffered too large uncertainties to be conclusive. In summary,
the binary scenario has not been disproved, but its ability to reproduce all the observations in GCs still has
to be demonstrated.

\begin{figure}
\begin{center}
\resizebox{\hsize}{!}{\includegraphics{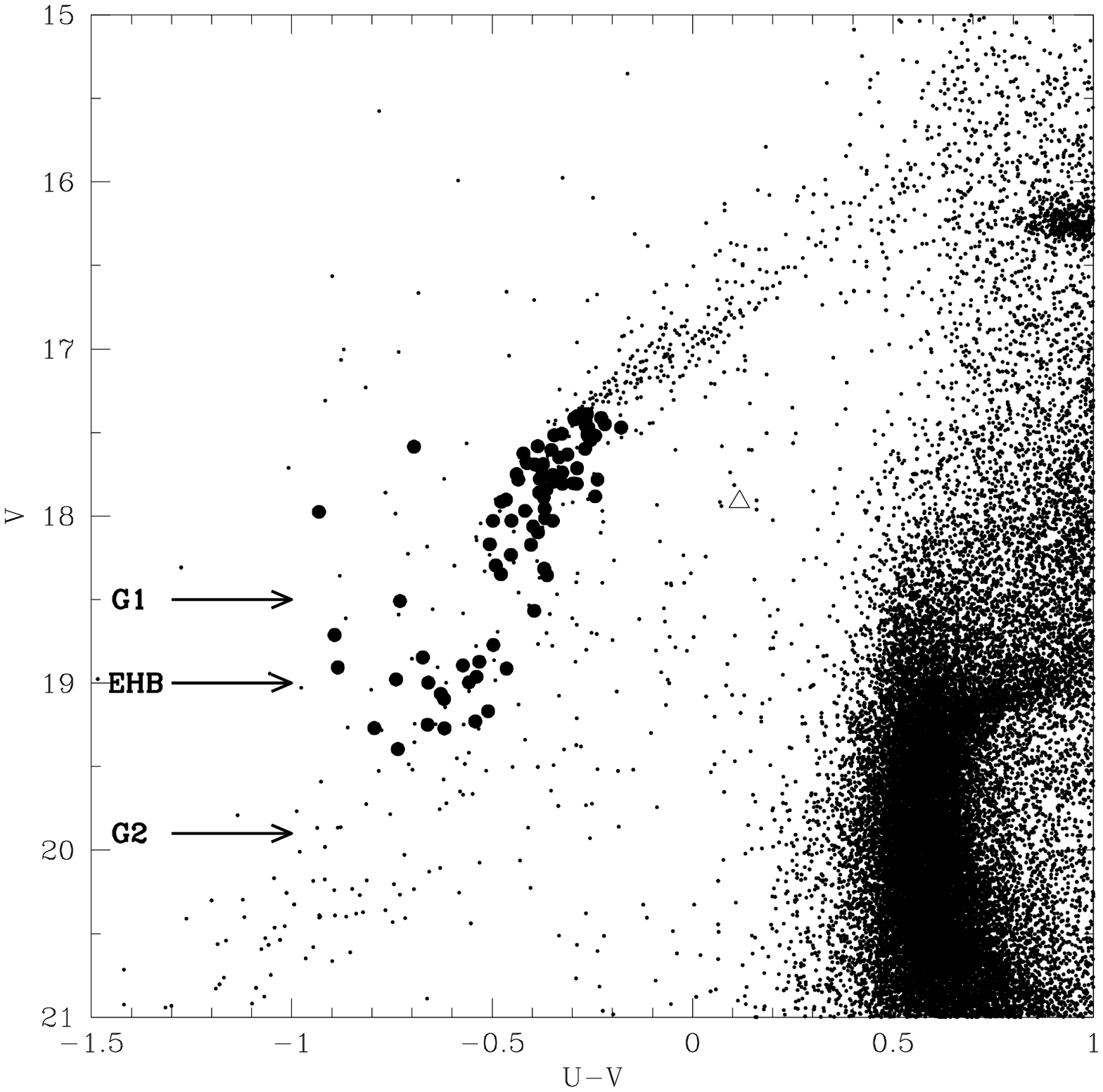}}
\resizebox{\hsize}{!}{\includegraphics{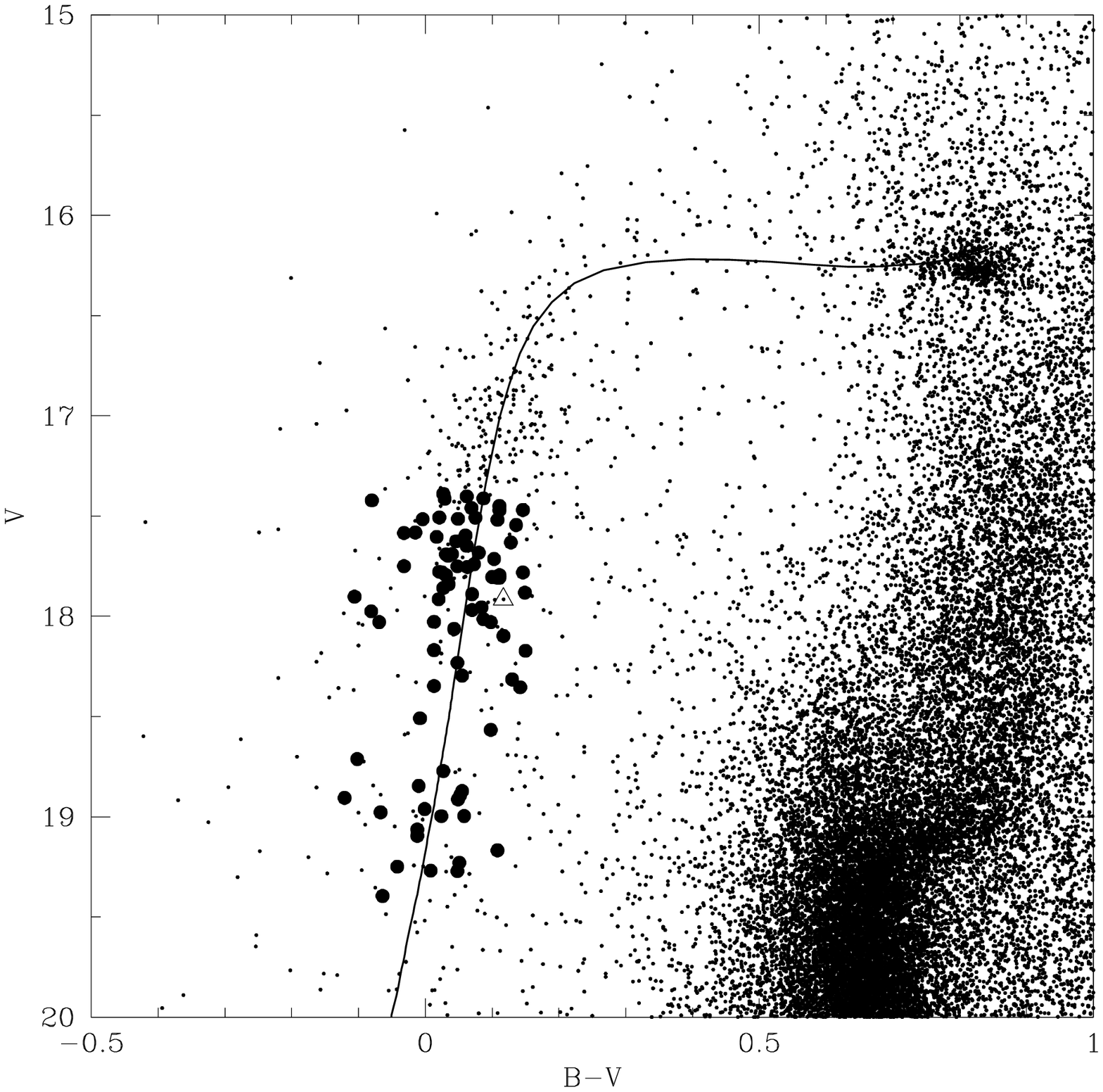}}
\caption{Position of the targets in the cluster color-magnitude diagrams. {\it Upper panel}: $V$ vs ($U-V$)
diagram, where the arrows indicate the location of the two gaps discussed in the text (G1 and G2), and
the magnitude corresponding to T=$20\,000$~K (the formal definition of the EHB). {\it Lower panel}: $V$ vs
($B-V$) diagram with superimposed the ZAHB model from \citet{Cassisi99} used to estimate stellar temperatures.
The star \#37345 discussed in the text is indicated with an empty triangle.}
\label{f_cmd}
\end{center}
\end{figure}

An alternative model of EHB star formation has received great attention in recent years: one incorporating
the primordial helium enhancement. A super-solar surface helium abundance was recognized early on as a possible
cause of the heavy-mass
loss underlying the formation of EHB stars, but non-canonical mixing phenomena had to be invoked
\citep{vonRudloff88,Denissenkov03,Sweigart79,Sweigart97}. The discovery of multiple stellar populations
in GCs \citep{Bedin04,Piotto07} opened a new frontier, because the observational results of \citet{Piotto05} 
apparently imply that the bluer main-sequence (MS) in \object{$\omega$ Cen} is most probably populated by
He-enriched stars. The currently preferred scenario is that these objects constitute a second stellar generation
that formed from material polluted by either intermediate-mass AGB stars \citep{DAntona02,Renzini08}, or rapidly
rotating massive MS stars \citep{Maeder06,Decressin07}. The models of multiple populations with different helium
abundances successfully reproduce both the MS splitting and the multimodal HB morphology of both
\object{$\omega$ Cen} \citep{Lee05} and \object{NGC\,2808} \citep{DAntona05}. The He-enhancement thus represents
a promising model, and alternative to the binary scenario, for the formation of EHB stars in GCs.

In this paper, we present our results of a search for EHB binaries in \object{NGC\,2808}. Preliminary results,
pointing to a close binary fraction higher than in \object{NGC\,6752}, were presented by \citet{Moni08b}
and \citet{Moni10}. In this context, this cluster is a key object because it is noticeably younger than
\object{NGC\,6752} \citep{DeAngeli05}, allowing the study of the $f_\mathrm{c}$-age relation foreseen by the
binary scenario. Moreover, this cluster represents one of the greatest successes of the He-enhancement scenario,
because \citet{DAntona05} and \citet{DOrazi10} were able to model both its multimodal HB \citep{Bedin00}
and the multiple MS \citep{Piotto07} with three stellar populations of increasing primordial helium abundance.


\section{Observations and data reduction}
\label{c_data}

We selected 83 hot HB stars in \object{NGC\,2808} from the photometric catalog of \citet{Momany03}, of magnitude
between $V$=17.4, corresponding approximatively to the Grundahl jump \citep{Grundahl99}, and the limiting
magnitude $V$=19.5 imposed by program feasibility. In Table~\ref{t_datatarg}, we give the IDs, the coordinates,
and the photometric data of the targets from \citet{Momany03}, and their location in the cluster CMDs shown in
Figure~\ref{f_cmd}.

\begin{figure}
\begin{center}
\resizebox{\hsize}{!}{\includegraphics{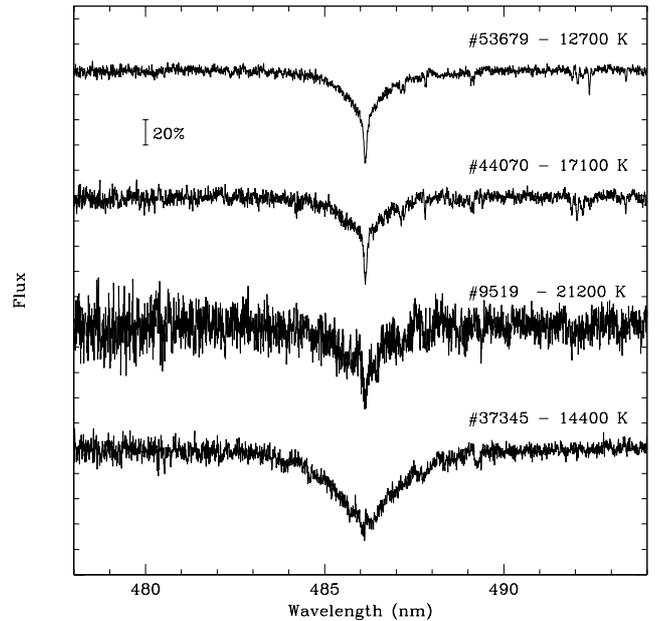}}
\caption{Normalized spectra of four target stars, obtained shifting all the collected spectra to laboratory
wavelength and summing them. The star ID and temperature is given. The spectra were vertically shifted to avoid
overlap.}
\label{f_spectra}
\end{center}
\end{figure}

Twenty-four single spectra of target stars were collected in fourteen epochs between January 11 and March 24, 2007,
with the GIRAFFE spectrograph at the VLT-UT2 telescope, in both visitor and service mode. The instrument setup H7A
provided high-resolution (R=18\,000) spectra centered on the H$_\beta$ line. The temporal sampling was carefully
planned to maximize the detection probability of binaries with any period up to 100~days. Exposures were
acquired in pairs to be later summed, except when bad weather conditions forced us to stop observations after
the first frame. We thus collected fourteen epochs of data over 2.5 months. The log of the observations is given
in Table~\ref{t_obs}, where each exposure is identified with a unique ID, and the epoch at the middle of the
acquisition period is indicated, along with the exposure time and the observing mode (v=visitor, s=service).

\begin{table}
\begin{center}
\caption{Log of the observations.}
\label{t_obs}
\begin{tabular}{c| c c c}
\hline
\hline
Exposure ID & Epoch & t$_\mathrm{exp}$ & Mode \\
 & JD$-$2450000 & s & \\
\hline
1-1 & 54111.26437 & 2x 3000 & v \\
1-2 & 54111.33528 & 2x 3000 & v \\
2-1 & 54112.25032 & 2x 3300 & v \\
2-2 & 54112.32817 & 2x 3300 & v \\
3-1 & 54114.24152 & 2x 3000 & v \\
3-2 & 54114.29434 & 1x 3000 & v \\
4-1 & 54116.25362 & 2x 3375 & v \\
4-2 & 54116.31284 & 1x 3375 & v \\
\hline
5-1 & 54127.16030 & 1x 1980 & s \\
5-2 & 54127.25886 & 2x 2770 & s \\
5-3 & 54127.33708 & 2x 2770 & s \\
\hline
6-1 & 54182.10276 & 2x 2770 & s \\
6-2 & 54182.16923 & 2x 2770 & s \\
6-3 & 54183.06863 & 1x 2770 & s \\
\hline
\end{tabular}
\end{center}
\end{table}

Data were reduced with the dedicated CPL-based pipeline available at the ESO web site. Because of the
extremely low signal collected for the hottest targets, we performed many trial reductions
to find the choices and parameter sets that maximized the output quality. The frames were de-biased and
flat-fielded with standard procedures based on the frames collected within the standard calibration plan. The
dark current was found to be non-negligible only along the top edge of the CCD, not used in our work, and no
dark correction was applied to avoid the corresponding decrease in S/N by 10-15\%. We gave particular
attention to the wavelength calibration (wlc), whose defects can easily affect the radial velocity (RV)
measurements. The goodness of the wlc was checked by analyzing the spectra of the lamp fibers acquired
simultaneously with target stars. This reduction step was particularly problematic, because we
found that running the complete wlc routine resulted in an incorrect solution, with a deviation from the correct
one that increased with wavelength and fiber number, up to 10-15~km~s$^{-1}$. We therefore adopted the standard
solution for the H7A setup, included in the instrumental package downloadable from the GIRAFFE web site,
allowing the pipeline to use the lamp fibers to find rigid shifts and changes in the spectral geometry on the
chip. After the final extraction, the lamp fibers showed only small random deviations from laboratory
wavelengths (0.3~km~s$^{-1}$ rms). This wlc error is small compared to uncertainties in the RV measurement,
and can be safely neglected in the final error budget. Finally, science spectra were extracted using both
an optimum algorithm \citep{Horne86} and a simple sum. We found that these two methods were in general
equivalent and the choice did not alter the results, but in some noisy spectra one or the other returned
more precise measurements. This was probably due to small cosmetic defects or noise spikes being treated
differently by the two algorithms. We therefore preferred optimum-extracted spectra, but we opted for a simple
sum in the few cases in which this clearly returned smaller RV errors. The background flux was estimated by
averaging nine fibers allocated to the sky and, after subtracting their mean spectrum from those of the targets,
we checked that the weak interstellar emission in the core of the H$_\beta$ line had been effectively removed.
The spectra were then trimmed to retain only the central region (4780-4930~\AA), and we normalized them fitting
a linear relation to the continuum on both sides of the H$_\beta$ line. We verified that a higher order
polynomial was not required in the normalization, as there was no appreciable change in either the fitted
function and or the results. As a final step of the reduction, the spectra forming a pair of exposures (see
Table~\ref{t_obs}) were added. Some example spectra are shown in Figure~\ref{f_spectra}, for two stars at the
edge of the temperature range (12\,700 and 21\,200~K) and one of intermediate temperature, plus the star
\#37345, discussed later. The presented spectra are the sum of all the spectra collected for each star, after
shifting them to laboratory wavelengths.

\begin{figure}
\begin{center}
\resizebox{\hsize}{!}{\includegraphics{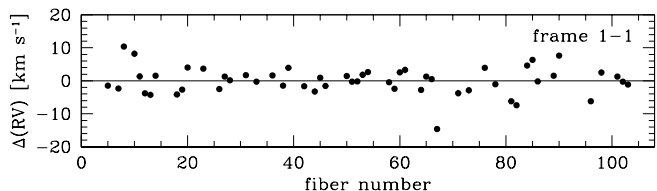}}
\caption{Difference between the RV measured in the first frame (frame 1-1 of Table~\ref{t_obs}) and the weighted
average for each star, plotted as a function of the fiber number. Only the 53 brightest stars, with measurement
uncertainty lower than 5~km~s$^{-1}$, are plotted.}
\label{f_syst}
\end{center}
\end{figure}

The observed HB was fitted with the zero-age HB model (ZAHB) of \citet{Cassisi99} with metallicity
[Fe/H]=$-$1.10 \citep{Carretta06} to derive a temperature scale along the HB. The procedure was not
straightforward using the ($U-V$) color, and uncertainties remained in the determination of the required
distance modulus and reddening. These problems could be due to the use of the $U$ band, because
\citet{DAlessandro2010} showed that, at shorter wavelengths, it is impossible to fit the HB of \object{NGC\,2808}
with one single population of fixed helium content. Therefore, the fit was performed in the $V$ versus
(hereafter vs.) ($B-V$) plane, where the closest match was found by assuming that (m-M)$_\mathrm{V}$=15.7 and
E($B-V$)=0.15, in good agreement with
\citet{Bedin00}. The fit is shown in the lower panel of Figure~\ref{f_cmd}. The temperature of each target was then
estimated from the point of the model ZAHB closest to the observed position. Varying (m-M)$_\mathrm{V}$ and E($B-V$)
between values that still gave a reasonably good fit, we estimated that the uncertainty in the temperature should be
on the order of 10\%. The temperature derived for each star is given in Table~\ref{t_datatarg}.

In the selected sample, there are eight stars hotter than 20\,000~K and, following the canonical definition, they
can be considered EHB stars. However, in the CMD of \object{NGC\,2808} there is no underpopulated region
in correspondence to this temperature as in, for example, \object{NGC\,6752} and \object{M80}, where the EHB is
separated from cooler HB stars. In contrast, \citet{Sosin97} identified two clear gaps in
\object{NGC\,2808} at about $V$=18.4 ($\sim$16\,500~K) and $V$=20 ($\sim$25\,000~K), called G1 and G2 respectively,
by \citet{Bedin00}. The cluster blue HB is thus divided into three sections, called EBT1
(T$_\mathrm{eff}\leq$16\,500~K), EBT2 (16\,000$\leq$T$_\mathrm{eff}\leq$25\,000~K), and EBT3
(T$_\mathrm{eff}\geq$25\,000~K). Following this scheme, the sample contains 62 EBT1 and 21 EBT2 stars, but no
star in the faint EBT3 group was observed.

We note that the temperature associated with the gaps was obtained by fitting a canonical model to
the observed HB, and the spectroscopic measurements of \citet{Moehler04} agree with this temperature scale.
However, \citet{DAlessandro2010} proposed a new scale, based on a multi-population model of three
stellar generations with increasing helium content. In their calculations, the G1 and G2 gaps approximatively
coincide with the canonical start (20\,000~K) and end (31\,000~K) of the EHB, which is thus separated from
cooler stars by an underpopulated region even in the CMD of this cluster.

\begin{figure}
\begin{center}
\resizebox{\hsize}{!}{\includegraphics{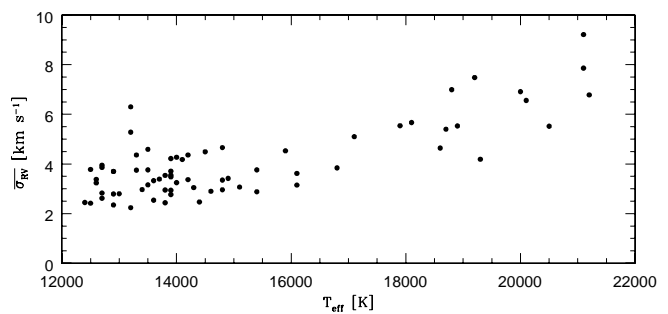}}
\caption{Mean error of each star as a function of effective temperature.}
\label{f_ertef}
\end{center}
\end{figure}


\section{RV measurements}
\label{c_measure}

The RVs were measured by means of the cross-correlation (CC) technique \citep{Tonry79} implemented in the
{\it fxcor} IRAF\footnote{IRAF is distributed by the National Optical Astronomy Observatories, which are
operated by the Association of Universities for Research in Astronomy, Inc., under cooperative agreement with
the National Science Foundation.} task. The CC was restricted to the spectral interval 4840-4880~\AA, i.e. the
H$_\beta$ line with its full wings. The template was extracted from the synthetic library of \citet{Munari05}.
Experiments with model spectra of different temperature, gravity, and metallicity showed that a change in these
parameters does not affect the results, while the errors are very sensitive to this choice, and the templates
with a narrower H$_\beta$ usually provide smaller errors for all the stars. These conclusions are also
supported by the detailed analysis of \citet{Morse91}. Variations in the shape of the template line are indeed
not expected to shift the center of the CC function (CCF), but can affect the resulting uncertainties. We also
found that the inclusion of weak metallic lines in the synthetic spectra enhanced the errors without
adding real information, because they were either absent or not visible in the noisy target spectra. Consequently,
the template finally adopted for all the stars was a synthetic profile of the H$_\beta$ line, obtained by
removing all weaker lines from a synthetic spectrum at 20\,000~K, $\log{(g)}$=5, and cluster metallicity. It
must be noted that the RV measurements are unaffected by any bias or uncertainty in the estimate of the target
temperature, because the choice of the template was independent of this temperature and, as discussed, this
choice did not alter the results.

\begin{figure}
\begin{center}
\resizebox{\hsize}{!}{\includegraphics{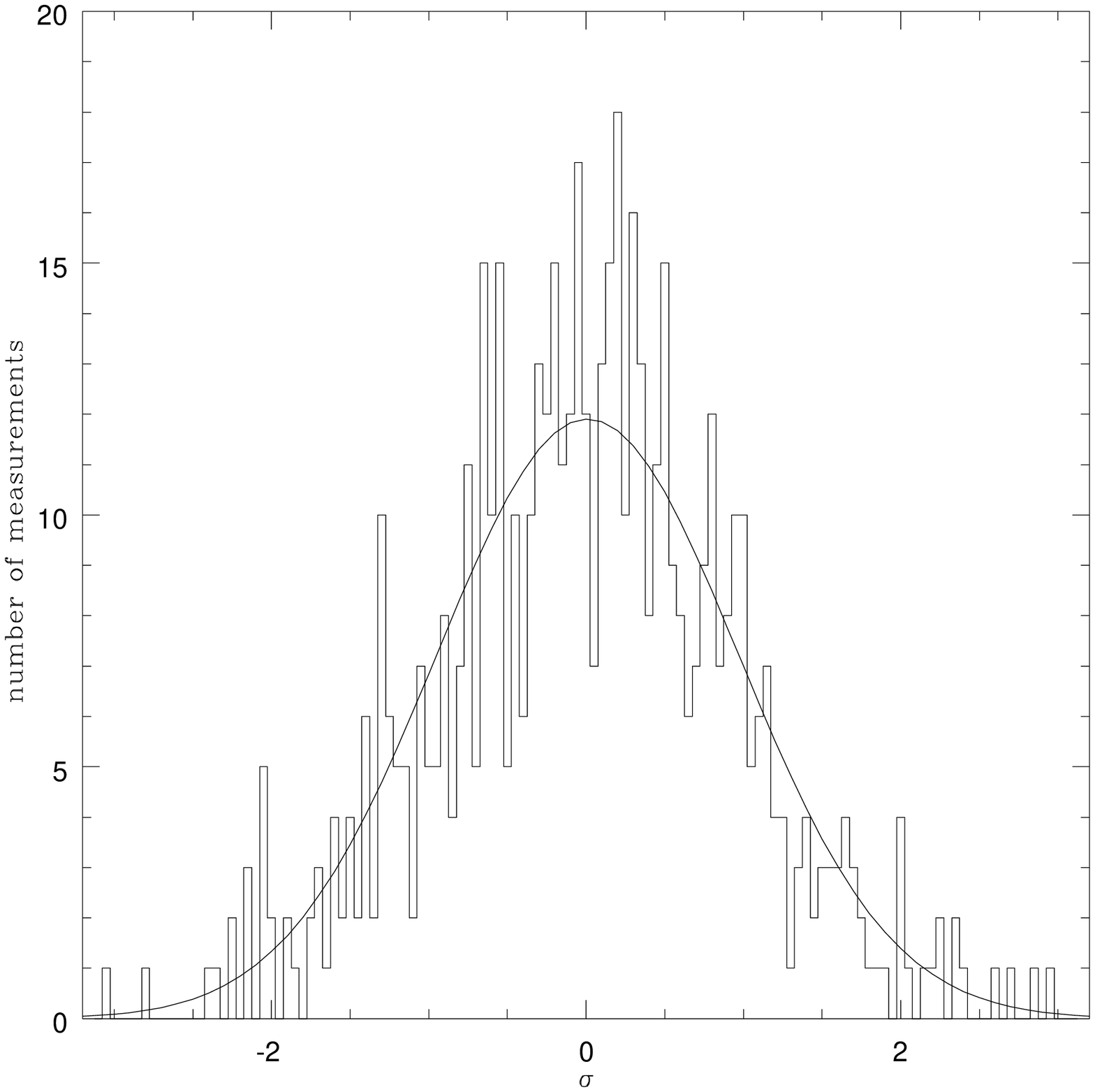}}
\resizebox{\hsize}{!}{\includegraphics{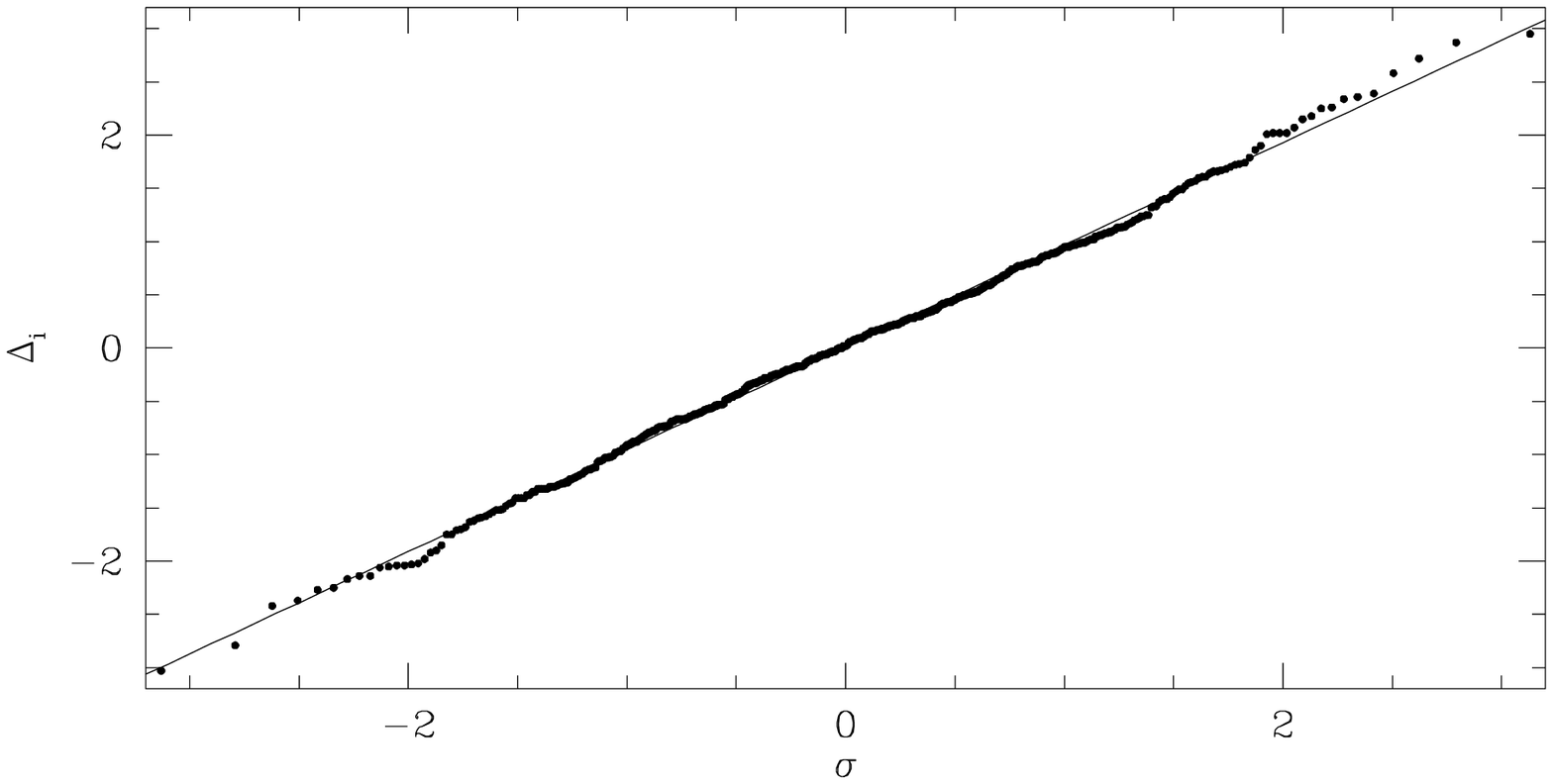}}
\caption{{\it Upper panel}: histogram of the distribution of the residuals with respect to the weighted average
of each star, in unit of the associated error (Equation~\ref{e_errors}). Overplotted to the histogram, a
Gaussian with $\sigma$=0.96 and centered in 0.01 is shown. {\it Lower panel}: probability plot of the same
residuals of the upper panel. The straight line, with intercept 0.01 and slope 0.96, indicates the least squares
fit of the data.}
\label{f_errors}
\end{center}
\end{figure}

The RV was determined by fitting the core of the CCF with a Gaussian profile. While the procedure was
straightforward for cool stars, it became problematic for the fainter ones, whose spectra were much noisier
down to signal-to-noise ratio (S/N)$\approx$3 in the worst cases. Measuring RVs at such low S/N with only one
wide line is a challenging task, and we with experimented different methods to improve the accuracy of the
measurements: the application of a narrow Fourier filter \citep{Brault71} sometimes helped, while the S/N of very
noisy spectra was often increased by degrading the resolution by a factor of 2-3 and then rebinning accordingly.
Although a lower resolution affects the precision that can be achieved,
the noise is by far the dominant source of uncertainty when the stellar flux is very weak. In some cases, we
used only one spectrum of a pair, because there was a great difference in the spectral quality of the two, and
the addition of the noisier one degraded the resulting measurement. We did not find a unique scheme for
returning the optimal results for all the spectra, but we employed in each case the procedures (filtering,
rebinning, use of a single spectrum) providing the best results in terms of shape and noise of the CCF, height of
its central peak, and goodness of its Gaussian fit. These procedures were designed only to reduce the errors, and
the results were stable when experimenting with different combinations of procedures and involved parameters,
varying by no more than a few~km~s$^{-1}$. When this was not the case, the measurement was judged unreliable and
excluded, as when the results were sensitive to changes in either continuum normalization or extraction algorithm.

\subsection{Systematic errors}
\label{c_systematics}

\begin{figure}
\begin{center}
\resizebox{\hsize}{!}{\includegraphics{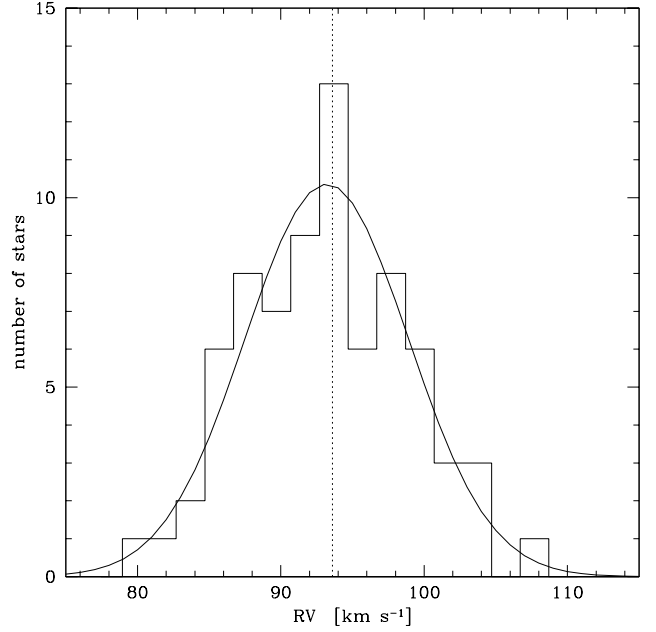}}
\caption{Distribution of the absolute RVs of the observed stars. The curve shows a Gaussian centered on the
mean value (93.2~km~s$^{-1}$) and $\sigma$ equal to the observed dispersion (5.7~km~s$^{-1}$). The dotted line
indicates the cluster RV from \citet{Harris96}.}
\label{f_absRV}
\end{center}
\end{figure}

After the correction to heliocentric RVs, we verified that no systematic error was present in the results. First,
we checked the zero-point of each frame by averaging the RVs of the 53 brightest stars, excluding measurements with
errors larger than 5~km~s$^{-1}$. We thus derived the corrections to reduce each frame to the same zero-point,
although they were lower than 1.5~km~s$^{-1}$, i.e. well within the typical error of the 53 stars
(3.5-4~km~s$^{-1}$). We then plotted, for each frame, the residual of each star with respect to its
weighted-averaged RV as a function of the fiber number, to check for the presence of a systematic effect that varied
with position on the CCD, as done by \citet[][Figure~5 and~7]{Moni06a}. The plot relative to the first frame
is shown in Figure~\ref{f_syst} as an example. The average value of these residuals
was always lower than 0.1~km~s$^{-1}$, indicating that any offset between exposures was correctly
removed in the previous step. Moreover, the linear and third-order fit never differed from zero by more
than 1~km~s$^{-1}$, proving that there is no residual trend in the measured RVs, and the results are free from
systematics well beyond the typical random errors of 3-4~km~s$^{-1}$.

\subsection{Errors}
\label{c_errors}

\begin{figure}
\begin{center}
\resizebox{\hsize}{!}{\includegraphics{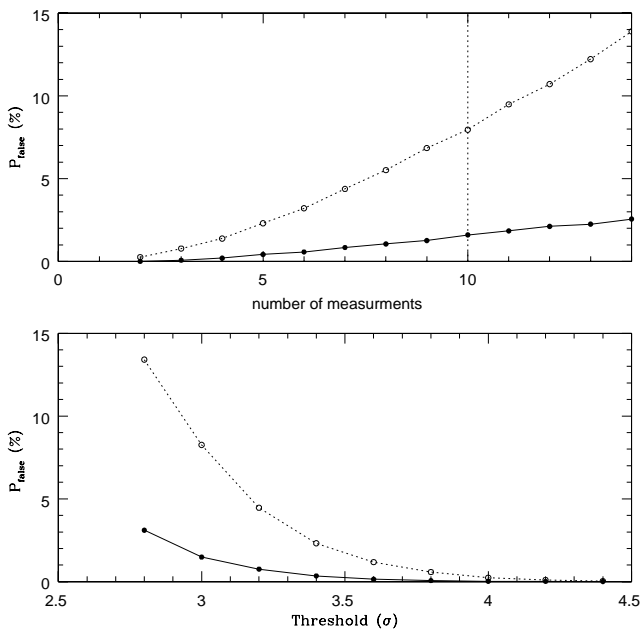}}
\caption{{\it Upper panel}: Probability of a false detection for the ''relative" criteria (empty dots and dotted
line) and the ''absoute" criteria (full dots and line) defined in the text, assuming a 3$\sigma$ threshold for
both. {\it Lower panel}: False detection probability for the same criteria, as a function of the threshold, for
ten measurements.}
\label{f_fdetprob}
\end{center}
\end{figure}

Any search for binaries by means of multi-epoch RV measurement is based on the comparison between the observed
variations and the uncertainties. A precise definition of the errors is therefore of fundamental importance
for the correct interpretation of the results.

The uncertainty associated with each measurement was defined as the quadratic sum of the error of the CC
technique \citep[as defined by][]{Tonry79}, the wlc error defined in Sect.\,\ref{c_data}, and the uncertainty
introduced when correcting the zero-point of each frame (Sect.\,\ref{c_systematics}). This last quantity was defined
as the rms of the corrections applied to each frame, i.e. the scatter in the zero-point of the frames around
the mean value. The CC error completely dominates the error budget, being typically 2-3~km~s$^{-1}$ for bright
stars and increasing with temperature up to 5-8~km~s$^{-1}$ for the faintest targets. Thus, the quadratic sum
of the wlc and zero-point errors (0.3 and 1.1 ~km~s$^{-1}$, respectively) makes a negligible difference for
all but the brightest targets. The mean error for each star is plotted in Figure~\ref{f_ertef} as a function
of effective temperature. The errors constantly increase with the temperature along the HB,
reflecting the decreasing S/N of the
collected spectra. The mean value of EBT1 targets (T$_\mathrm{eff}\leq$17\,000~K) ranges from 2.5 to
$\sim$4.5~km~s$^{-1}$, with the exception of a couple of deviant stars, with an average value of 3.5~km~s$^{-1}$.
Hotter stars show a higher scatter and a steeper gradient, with a mean value of 6.0~km~s$^{-1}$.

To test the reliability of the errors, we analyzed the distribution of the residuals ($\Delta_i$) of the i-th
measurement (RV$_i$) with respect to the weighted average ($\overline{\mathrm RV}$), in units
of the associated error $\sigma_\mathrm{RV,i}$
\begin{equation}
\Delta_i=\frac{RV_i-\overline{\mathrm RV}}{\sigma_\mathrm{RV,i}}.
\label{e_errors}
\end{equation}
If the observational uncertainties are the only cause of the variations, and the errors $\sigma_\mathrm{RV,i}$
represent them well, the residuals would follow a normal distribution centered on zero and with unit dispersion.
This analysis can easily be spoiled by RV variable stars, which add large values to the wings of the distribution.
We therefore excluded all the stars showing one or more residuals $\Delta_i\geq3.5$, as we later demonstrate that
no variation above this threshold should be expected from random errors alone (Sect.\,\ref{c_threshold}).
The resulting distribution of $\Delta_i$ is shown in the upper panel of Figure~\ref{f_errors}. Both the mean
value (0.01) and the standard deviation (0.96) confirm that the errors are indeed well defined, as they account
exactly for the random variations. The probability plot \citep{Lutz92} of these data reveals that they follow a
normal distribution, as they are aligned along a straight line of slope 0.96 and intercept 0.01. This would have
not happened if, for example, the exclusion of stars with $\Delta_i\geq3.5$ had produced a too narrow cut of
the wings of the distribution. In conclusion, the analysis reveals that the errors are reliable, representing well
the random uncertainties, and that they can safely be used in a statistical analysis.

\subsection{Absolute RVs and cluster membership}
\label{c_absRV}

The absolute RVs were calculated by weight-averaging all the measurements for each star, and the results are given
in column~7 of Table~\ref{t_datatarg}. The histogram of the distribution is shown in Figure~\ref{f_absRV}. The
mean value is $\overline{\mathrm RV}_\mathrm{abs}$=93.2~km~s$^{-1}$, in excellent agreement with
\citet[][February 2003 Web version, 93.6~km~s$^{-1}$]{Harris96}, and the observed dispersion is 5.7~km~s$^{-1}$.
In Figure~\ref{f_absRV}, we also overplot a Gaussian curve centered on $\overline{\mathrm RV}_\mathrm{abs}$
with $\sigma$=5.7~km~s$^{-1}$. The histogram is well described by a normal distribution, with no
deviating points, hence all the targets can be considered {\it bona fide} cluster members. The only exception is
the star \#37345, whose spectrum is shown in Figure~\ref{f_spectra}. This object was immediately recognized as
peculiar, because of its very broad H$_\beta$ line: rapidly rotating HB stars hotter than $\sim$11\,500~K have never
been found \citep{RecioBlanco02,Behr03}, while they are very common among hot MS stars. We note that this object
is much redder than the other HB stars in terms of $U-V$ color (of Figure~\ref{f_cmd}), but merges with
the HB population in the $V$-$(B-V)$ diagram (lower panel of the same Figure). The RV measurements was highly
uncertain because, as a consequence of the line broadening, the fit of the wide peak in the CCF was problematic.
When it converged, we derived RV$\approx$150~km~s$^{-1}$, very different from the cluster value. We therefore
conclude that the star \#37345 is most probably a background field MS B star.


\section{Analysis}
\label{c_results}

Eleven stars were not considered in our analysis, because their spectra were either too noisy for reliable
measurements or strongly contaminated by the emission lines of nearby lamp fibers. All these targets have no data
in the last two columns of Table~\ref{t_datatarg}.
The majority of targets were not allocated the same fiber in both GIRAFFE plates, as these are not equivalent
because of broken fibers. As a consequence, the spectra of six stars were free of lamp contamination when observed
with one plate, while their spectra collected with the other plate were damaged. Moreover, star \#45980 had no fiber
allocated in one plate. These seven targets were studied as the others, but were excluded from our statistical
analysis because, with only half of the spectra with respect to the other stars, their temporal sampling is very
different. Finally, one target was found to be a foreground star (Sect.\,\ref{c_absRV}). In conclusion,
with nineteen stars lost or excluded, the resulting sample comprises 50 EBT1 and 14 EBT2 stars, out of which six
EHB targets have T$_\mathrm{eff}\geq$20\,000~K.

The measurements for the four spectra that were not the sum of two exposures, i.e. the frames 3-2, 4-2, 5-1, and
6-3 (Table~\ref{t_obs}), were in general less reliable, and they were often excluded for faint stars. 
In the upper panel of Figure~\ref{f_detprob}, we plot the decrease in the binary detection efficiency, as a function
of period, if these four frames are not considered. Their exclusion clearly does not affect the survey much, because
the probability of detecting a true binary in general decreases by less than 3\%, except for two sensitive
periodicities (1 day and $\sim$18~days) where the loss is about 10-15\%. Therefore, we finally excluded them from the
statistical analysis. The measurements for these lower-quality frames were indeed always more uncertain, and were
often excluded anyway, affecting the uniformity of the measurements. We then consider only the remaining ten
epochs for each star.

\subsection{Detection threshold and detection probability}
\label{c_threshold}

The identification of binary candidates requires criteria to define when the observed variations can be
considered significant. The criteria must satisfy two desiderata: the probability of a false detection
(P$_\mathrm{false}$) must be negligible, while the probability of detecting a true binary (P$_\mathrm{det}$) must
be as high as possible, within the limitations imposed by the temporal sampling and the observational errors.
Usually the first point is satisfied if the statistical expectation is less than one false detection in the whole
survey, which in our samples of 64 targets means P$_\mathrm{false}\leq$0.014. A compromise between these
requirements is often needed, because more stringent criteria reducing P$_\mathrm{false}$ also reduce the
efficiency of detecting genuine RV variables.

\begin{figure}
\begin{center}
\resizebox{\hsize}{!}{\includegraphics{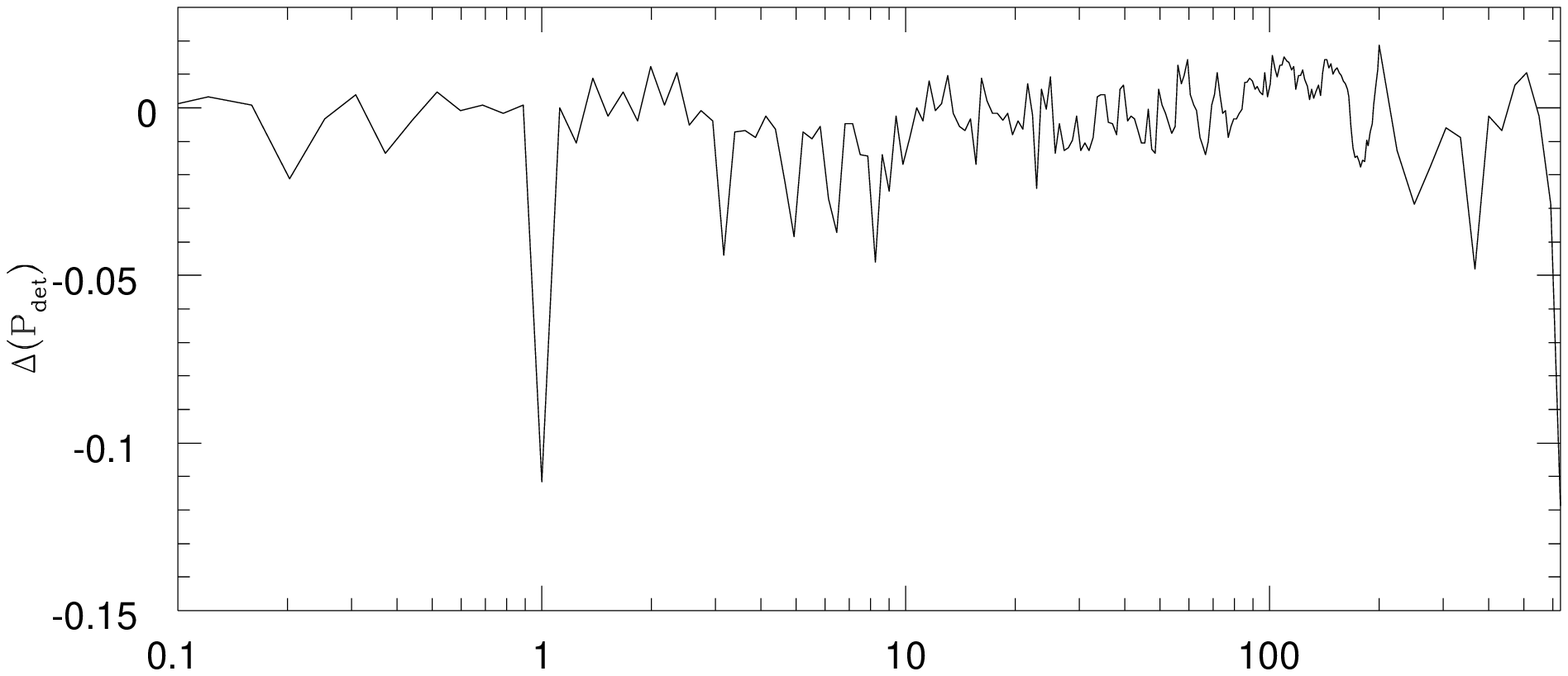}}
\resizebox{\hsize}{!}{\includegraphics{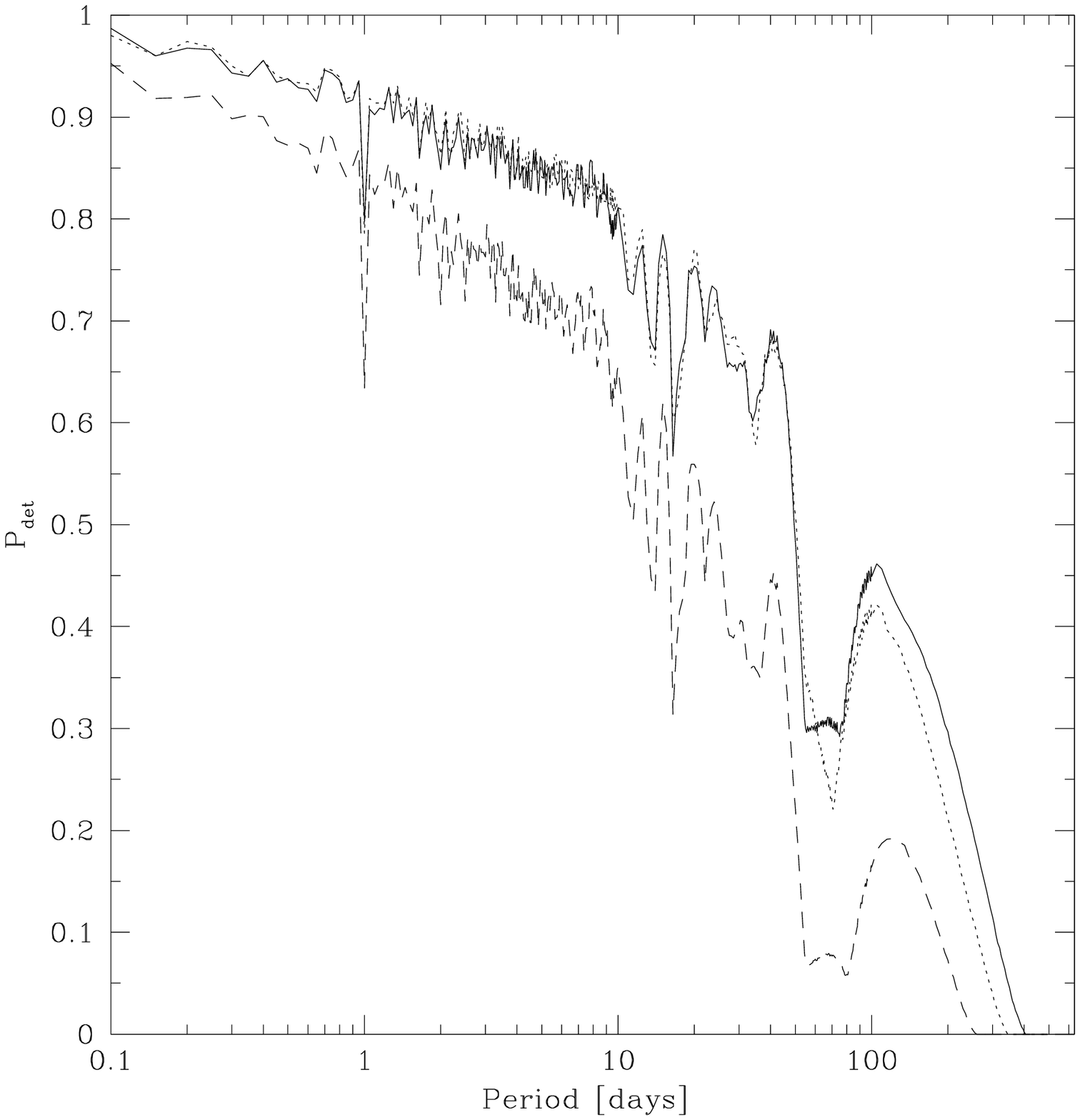}}
\caption{{\it Upper panel}: difference of the binary detection probability, as a function of period, when
including or not the four spectra which are not the sum of two exposures (frames 3-2, 4-2, 5-1, and
6-3, see Table~\ref{t_obs}). For the calculation we assumed an ''absolute" criteria with threshold
3.2$\sigma$, the same adopted in the present work, and a mean observational error of 3.5~km~s$^{-1}$.
{\it Lower panel}: detection probability as a function of period, calculated as described in the text. The full
and dotted lines indicate the results for the ''absolute" criteria with 3.2$\sigma$ threshold and the ''relative"
one with 3.8$\sigma$ criteria, respectively, assuming the typical mean error (3.5~km~s$^{-1}$) of stars cooler
than the G1 gap at $\sim$17\,000~K. The results for the ''absolute" criteria assuming a mean error of 6.0
km s$^{-1}$, characteristic of hotter stars, are shown with a dashed line.}
\label{f_detprob}
\end{center}
\end{figure}

The parameter P$_\mathrm{det}$ was estimated as a function of orbital period, generating 2500 synthetic binaries in
circular orbits of period $\wp$, comprising two stars of 0.5~M$_\odot$, uniformly distributed in the sin($i$)-$\theta$
space, where $i$ is the angle of inclination of the orbital plane and $\theta$ is the orbital phase. These systems were
then ''observed" with the same temporal sampling of our survey, and each star satisfying the criteria under
analysis represented a detection. The fraction of detections over the whole sample thus indicated the efficiency
of the survey for systems of period $\wp$. P$_\mathrm{false}$ was calculated by simulating 100\,000
sets of $N$ measurements drawn from a normal distribution centered on zero and of unity dispersion. Each set
represented the observations of a star with constant RV, affected by only random errors normalized to unity, and
$N$=10 for our survey. A false detection was claimed for each set of measurements satisfying the criteria for
binary detection, and P$_\mathrm{false}$ was estimated as the fraction of false detections over the whole sample
of 100\,000 attempts.

\begin{figure}
\begin{center}
\resizebox{\hsize}{!}{\includegraphics{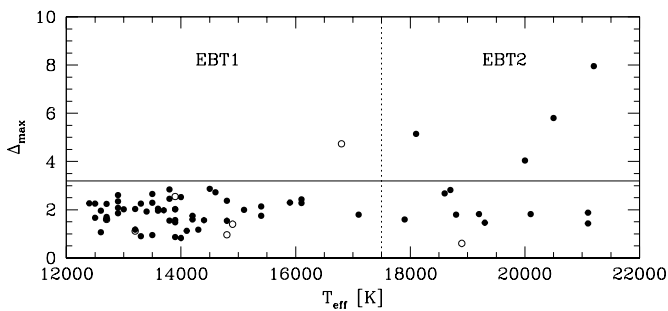}}
\caption{Maximum residual in units of the error (Equation~\ref{e_errors}) for each star, as a
function of temperature. The horizontal line indicates the adopted threshold for binary detection. Empty dots
represents stars with only partial temporal coverage, that are excluded from the statistical analysis.
The vertical dotted line indicates the approximate position of the G1 gap.}
\label{f_restef}
\end{center}
\end{figure}

In previous investigations of HB stars in GCs, the criteria adopted for the detection of a binary candidate was
that the measured RV variation was larger than 3$\sigma$, where $\sigma$ was either the error in the 
spectral shift among two epochs
\citep{Moni06a,Moni09}, or the quadratic sum of the errors in RV measurements \citep{Moni08a}.
This strategy worked well for surveys based on 4-5 epochs, but is unsuitable for our work: the upper
panel of Figure~\ref{f_fdetprob} shows that the probability of a false detection rapidly increases with the
number of measurements, and in our 10-epochs survey we should expect 8\% of non-binary targets ($\sim$6
stars) to violate the threshold due to random errors only. The method could be applied even here, but
to reduce the expected false detections to less than one star we should increase the threshold up to 3.8$\sigma$,
as indicated by the lower panel of Figure~\ref{f_fdetprob}, i.e. $\sim$25-30~km~s$^{-1}$ for the hotter stars.
In the same Figure, we also show P$_\mathrm{false}$ for an alternative criteria, which we define as ''absolute"
(while the older one is called ''relative"): a star is flagged as binary candidate if one (or more)
measurement differs by more than a certain threshold, in units of its associated error, from the weighted average.
The plot reveals that the reliability of this criteria is poorly affected by the higher number of measurements,
and that in our survey the threshold must be fixed to 3.2$\sigma$ to have P$_\mathrm{false} \leq$0.7\%,
i.e. less than 0.5 false detections expected among our 69 targets. In the present work, we adopt both this
criteria and this threshold for the detection of binaries. However, the results are completely equivalent to those
for a ''relative" criteria with threshold of 3.8$\sigma$, because P$_\mathrm{false}$ is the same, and P$_\mathrm{det}$
is very similar to within a few percent at all periods (see lower panel of Figure~\ref{f_detprob}).

The efficiency of the survey resulting from the adopted criteria, i.e. the probability of detection as a function
of the orbital period, is shown if Figure~\ref{f_detprob}. We considered two possibilities: a mean error of
3.5~km~s$^{-1}$, typical of the targets brighter than the G1 gap (T$_\mathrm{eff}\leq$17\,000~K), and 6.0
km s$^{-1}$ for the hotter EBT2 stars. In the first case, the survey can detect binaries with periods up to 400
days and, although the probability of detection drops below 50\% for periods longer than 50~days, still more than
one-third of the binaries with period 50-200~days can be found with our observations. For hotter stars, the larger
errors decrease at all periods the efficiency, which becomes negligible ($\leq$20\%) for P$\geq$50
days. In any case, the presence of HB binaries with periods longer than 10~days can be tested for the first time in
a GC, and these wide systems are poorly studied even among field sdB stars.

\subsection{Detected binaries}
\label{c_binaries}

\begin{figure}
\begin{center}
\resizebox{\hsize}{!}{\includegraphics{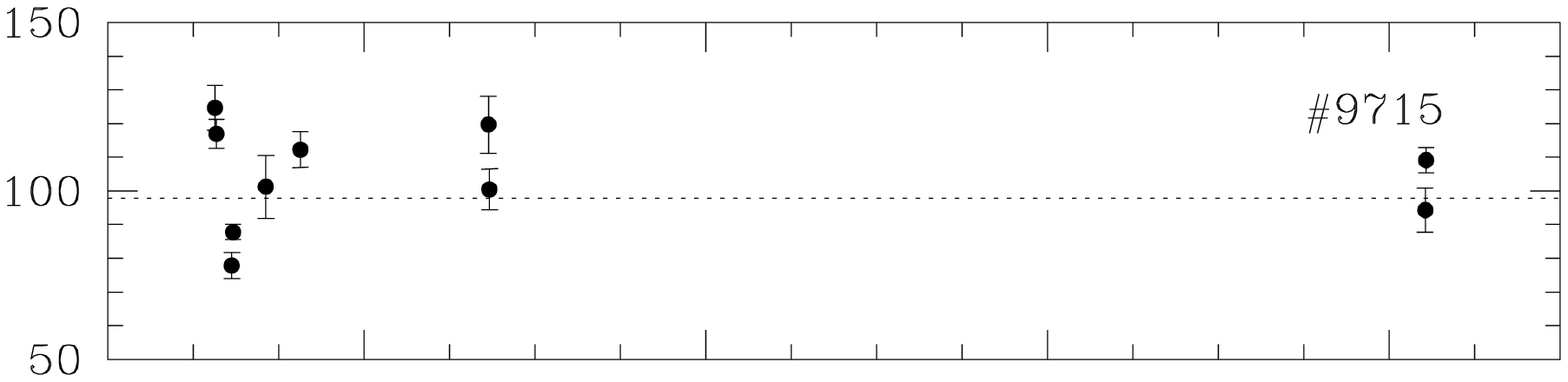}}
\resizebox{\hsize}{!}{\includegraphics{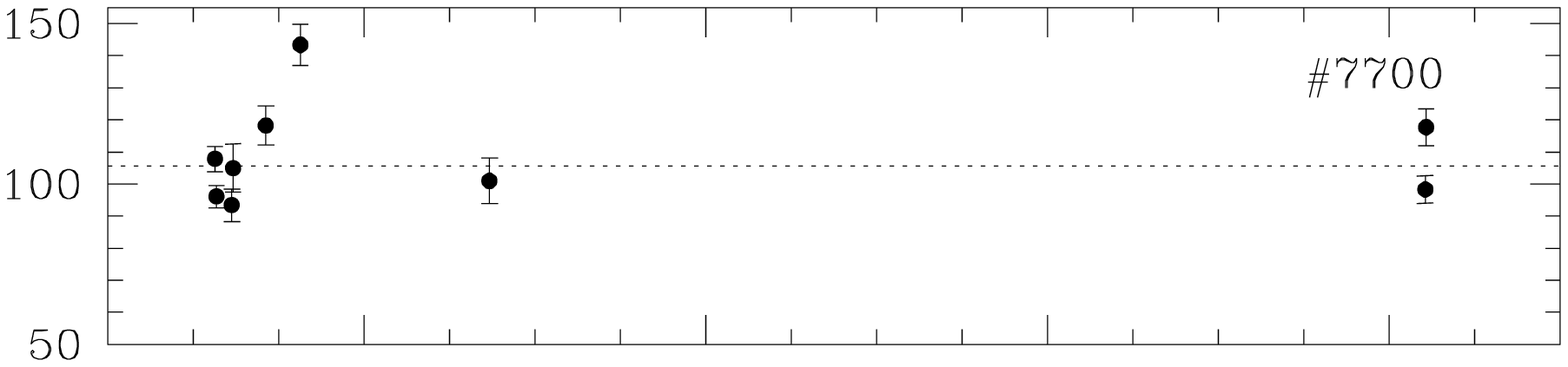}}
\resizebox{\hsize}{!}{\includegraphics{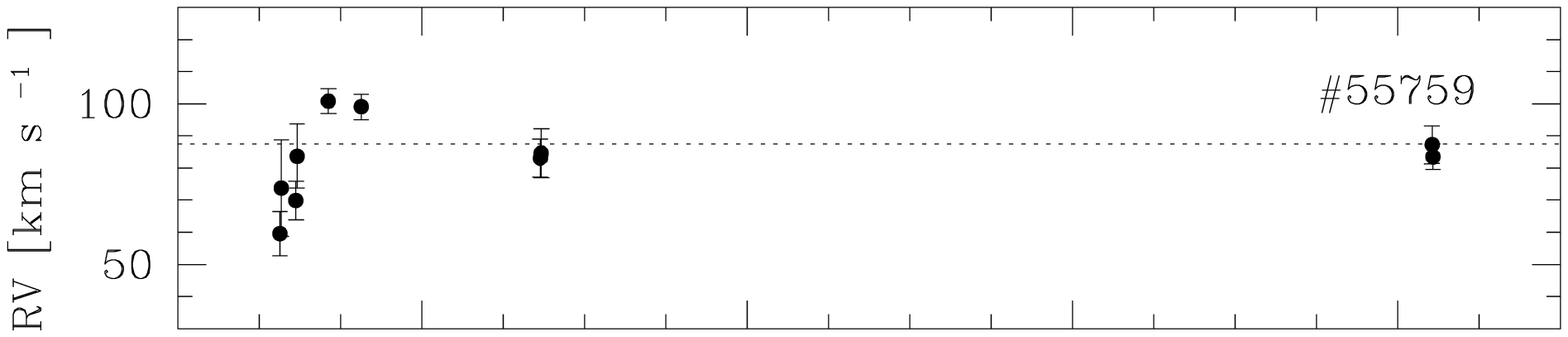}}
\resizebox{\hsize}{!}{\includegraphics{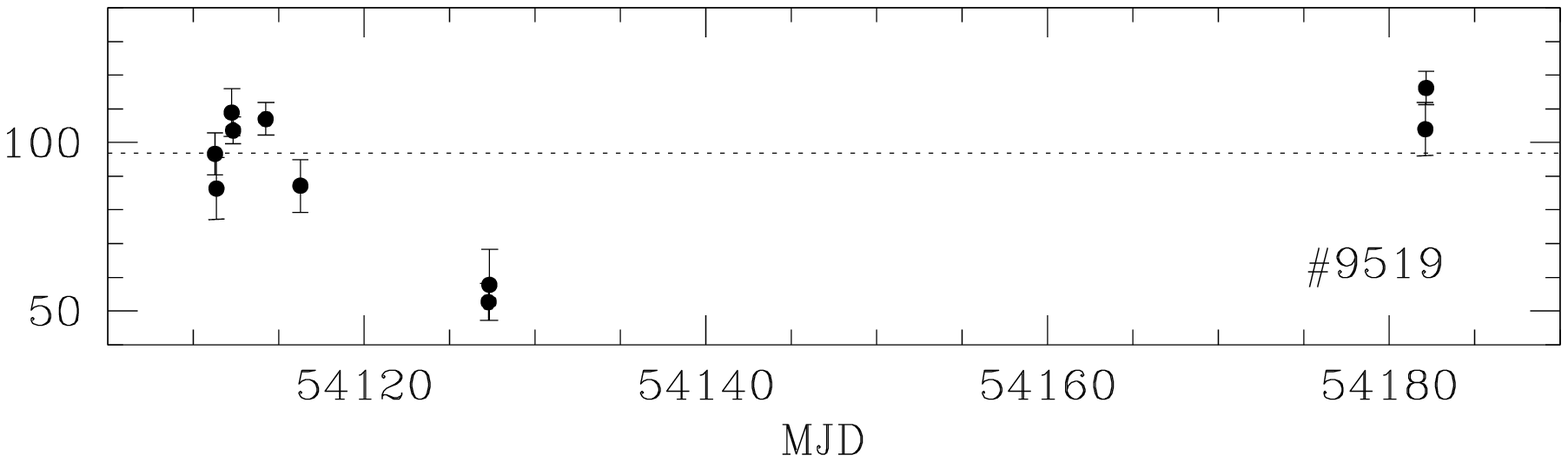}}
\caption{RV measurements of the four binary candidates. The dotted line indicates the weighted mean.}
\label{f_rvcurves}
\end{center}
\end{figure}

The results of the search for RV variable stars are summarized in Figure~\ref{f_restef}, where for each star we
plot the maximum residual $\Delta_\mathrm{max}$ defined in Equation~\ref{e_errors}. This value is also given in
column~8 of Table~\ref{t_datatarg}. Five targets show one or more significant deviations from the weighted mean,
and they must be considered candidate binaries. The cooler
one (star \#32670) is one of the seven targets with only partial temporal coverage mentioned at the beginning of this
section, and is excluded from the statistical analysis. The other four binary candidates are located in the
hotter half of the temperature range, and are all EBT2 stars fainter than the G1 gap. All but one of them
(\#9715) are hotter than 20\,000~K, thus fall within the canonical definition of EHB stars.
Their measured RVs are shown in Figure~\ref{f_rvcurves}.

We note that, although the hotter stars were excluded from the error analysis of Sect.\,\ref{c_errors}, the
stars with no variation above the threshold show no gradient in $\Delta_\mathrm{max}$ with temperature,
indicating that the errors are well-defined across the whole range. Moreover, there is a gap in the distribution of
$\Delta_\mathrm{max}$ between these stars and the five binary candidates, which reinforces the idea that they are
outliers. This confirms that the adopted threshold is indeed a good separation between statistical random errors
and real RV variations.

\subsection{Period estimate}
\label{c_period}

Our RV measurements are too few in number, and affected by too large errors, to attempt any estimate of the orbital
periods of the binary candidates. However, our survey can detect systems with periods up to 400~days
(Sect.\,\ref{c_threshold}), while previous investigations of GCs and the most extensive surveys among field stars
were limited to close binaries with periods shorter than 10~days. This implies that our results are not directly
comparable to those works without a minimum classification of the discovered systems. Hereafter, we define {\it close
binaries} as systems with $\wp\leq$10~days. They have been the common targets of previous investigations, and should
have undergone one or two CE phases. We also define {\it intermediate-period binaries} as objects with
10$\leq\wp\leq$400~days. We thus leave the definition of {\it wide binaries} to those with periods longer than
400~days, which is the typical product of RLOF \citep{Han02}, a kind of system undetectable in our survey.

\begin{figure}
\begin{center}
\resizebox{\hsize}{!}{\includegraphics{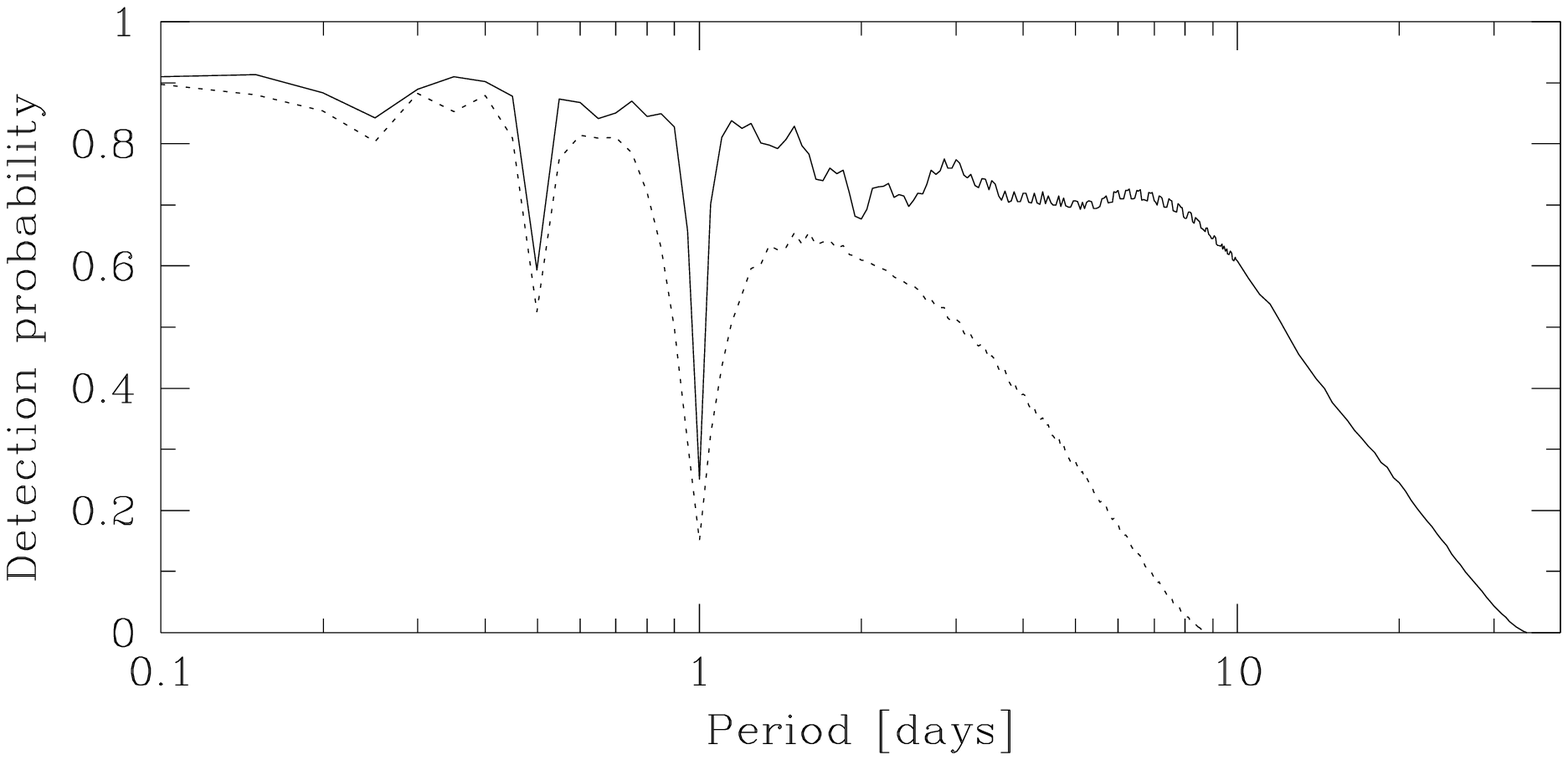}}
\resizebox{\hsize}{!}{\includegraphics{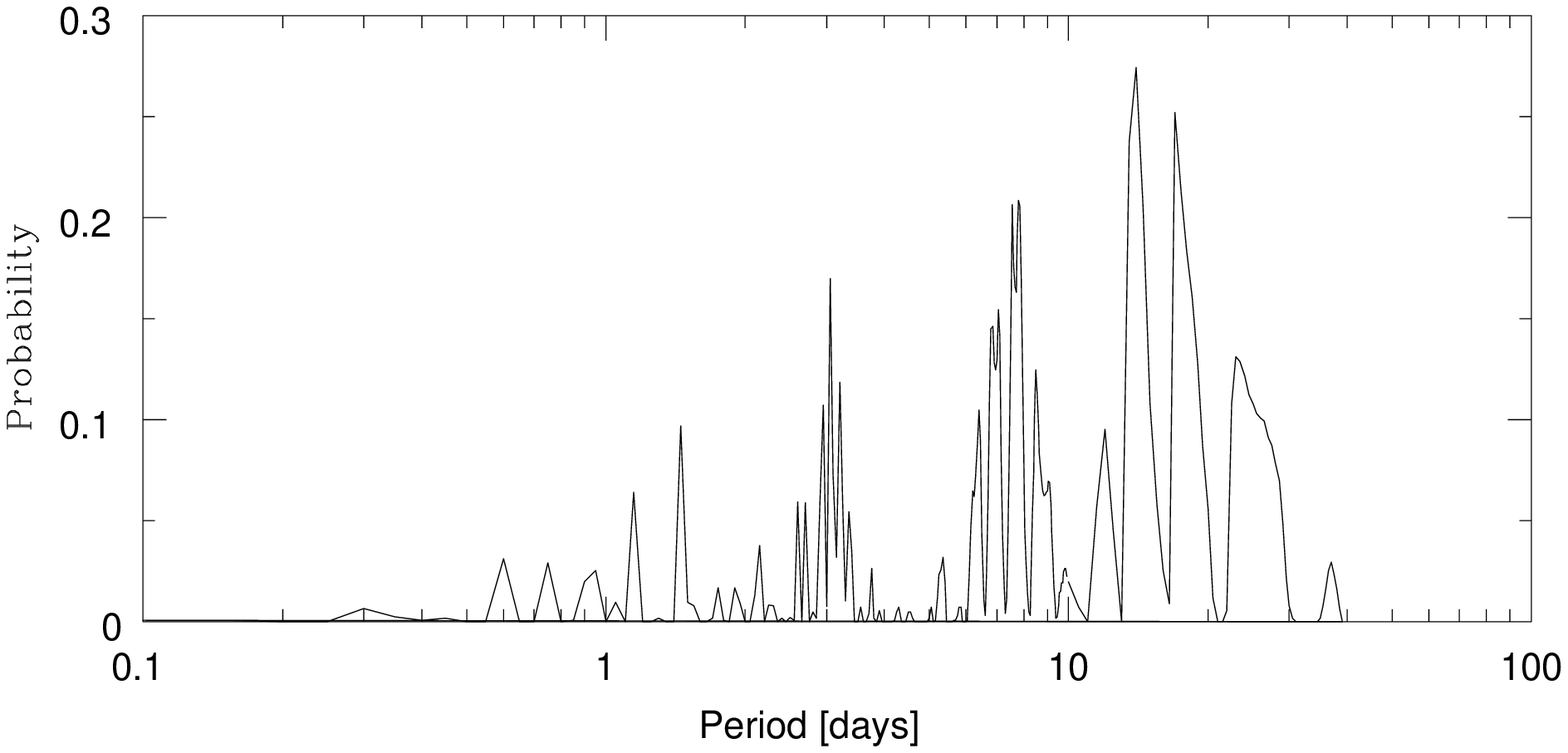}}
\caption{{\it Upper panel}: detection probability as a function of period, calculated as described in the text,
consideting only the first four (full line) or two (dotted line) nights. {\it Lower panel}: probability, as a
function of period, that an EHB binary would show a significant RV variation only in the fourth night of our
observations, as found for the star \#7700.}
\label{f_4nights}
\end{center}
\end{figure}

To help classify the binary candidates, we can rely on the adopted temporal sampling: the observations of the first
four closely-spaced nights were designed to detect short-period systems, while the later fifth and sixth epochs
were planned to help us discover longer periodicities. If our observations had been restricted to the first two nights
only, the stars \#9715 and \#55759 would still have been detected. As can be seen in the upper panel of
Figure~\ref{f_4nights}, systems with periods longer than 5~days would hardly have been found in such a two-night survey,
and all binaries with $\wp\geq$10~days would have passed completely undetected. The period of these two binary stars
must therefore be shorter than 10~days, and most probably even shorter than 5~days. These objects can thus safely be
classified as close binaries. Figure~\ref{f_4nights} also reveals that $\sim$80\% of the close binaries should have
been detected during the first four nights. The remaining 20\% mainly consists of nearly face-on systems showing
maximum RV variation near or below the detection threshold, as indicated by the inclusion of the other two
epochs not noticeably increasing the survey efficiency (compare Figs.~\ref{f_4nights} and~\ref{f_detprob}). The
star \#9519 shows variations above the threshold, but not in the first four nights, and therefore can safely be
classified as an intermediate-period binary of period longer than 10~days. The classification of the last binary
candidate (star \#7700) is less straightforward, because it passed undetected during the first three nights but it
had a significantly different RV during the fourth one. To clarify its nature, we calculated the probability that
a system of a given period behave as observed, i.e. the fraction of the synthetic binaries defined in
Sect.\,\ref{c_threshold} that would show a significant variation in the fourth epoch only. The results are plotted in
the lower panel of Figure~\ref{f_4nights}. The period is most likely between 5 and 20~days, although shorter values
are not impossible, and the most probable period is $\sim$15~days. Taking into account that, as argued by
\citet{Moni08a}, restricting the definition of close binaries to $\wp\leq$5~days would not make a great difference,
we can classify this object as an intermediate-period binary.

\subsection{Binary fraction}
\label{c_frac}

The probability of detecting $N_B$ binaries in a sample of $N$ stars is given by the expression
\begin{equation}
\mathrm{p} = \frac{\mathrm{N!}}{(\mathrm{N-N_{\mathrm{B}}})!\mathrm{N_{\mathrm B}}!}
(\mathrm{\bar{d}}f)^\mathrm{N_{\mathrm B}}(1-\mathrm{\bar{d}}f)^{\mathrm{N-N_{\mathrm B}}},
\label{eq_frac}
\end{equation}
where $f$ is the underlying binary fraction and $\bar{d}$ is the average probability of detection weighted by the
period distribution. This equation can also be interpreted as the probability that the (unknown) binary fraction is
$f$, when $N_B$ systems out of $N$ targets were detected. The calculation requires information about the underlying
period distribution, which is unknown. A Gaussian distribution in $\log{(\wp)}$ centered on zero has been
proposed for close field sdB systems \citep[e.g.][]{Maxted01,Napiwotzki04}, while wider systems are still poorly
studied. We assume a flat distribution at all periods, because \citet{Moni08a} demonstrated that the results
do not change by more than a few percent if other reasonable, but not proven, assumptions are made.

The binary fraction will be estimated separately for the EBT1 and EBT2 sections of the HB, which have different
samplings (Figure~\ref{f_cmd}) and detection probabilities (Figure~\ref{f_detprob}). These two groups of stars are
separated by an underpopulated region in the CMD, and could represent different stellar populations
\citep{DAntona05}. Stars hotter than 20\,000~K will be considered as EHB stars, a sub-group of the EBT2, although
this could be only a formal distinction since no known discontinuity separates them from the other EBT2 stars in the
CMD. It must also be recalled that this definition strongly depends on the model fixing the temperature scale along
the HB, and the multi-population model of \citet{DAlessandro2010} EBT2 and EHB actually coinciding. We also
distinguish between the fraction of close and intermediate-period binaries ($f_\mathrm{c}$ and $f_\mathrm{ip}$,
respectively). The analysis of $f_\mathrm{ip}$ is restricted to $\wp$=10-200~days in the EBT1 and to $\wp$=10-50~days
in the EBT2 and EHB, because the efficiency of the survey rapidly decays for longer periods. In the first case, this
choice does not affect the results, because no binary was detected and therefore the solution $N_B$=0 can be
applied to any period interval of interest. In contrast, this restriction could lead to wrong results for the
EBT2/EHB because, if one of the two candidate intermediate-period binaries had $\wp\geq$50~days, $f_\mathrm{ip}$
would be overestimated. However, we have already demonstrated that the star \#7700 has $\wp\leq$30~days (see
Figure~\ref{f_4nights}), and a similar analysis for the star \#9519 suggests that the probability that
$\wp\geq$50~days is negligible even for this star.

\subsubsection{EBT1}
\label{c_fracEBT1}

\begin{figure}
\begin{center}
\includegraphics[width=7.5cm]{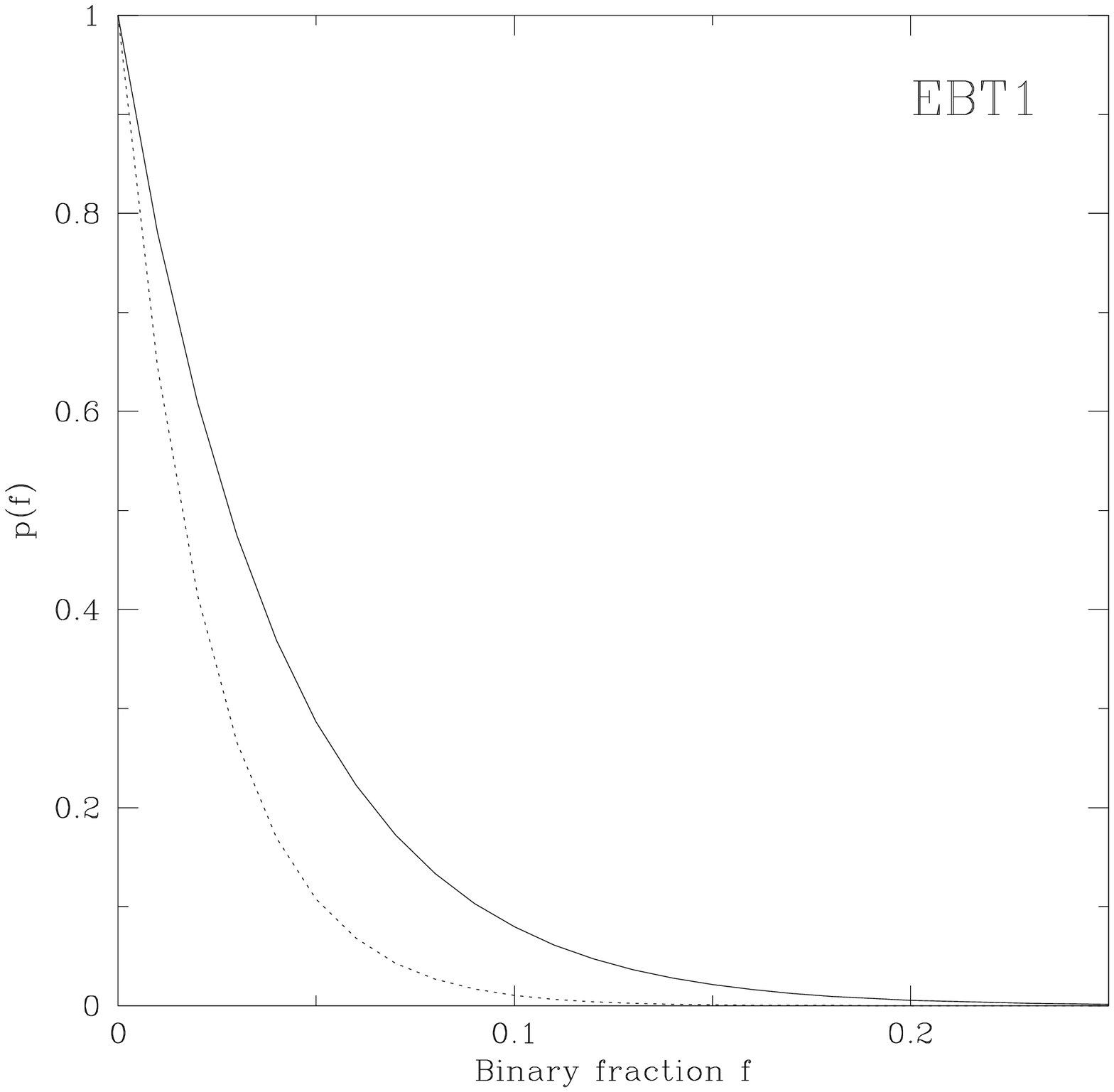}
\includegraphics[width=7.5cm]{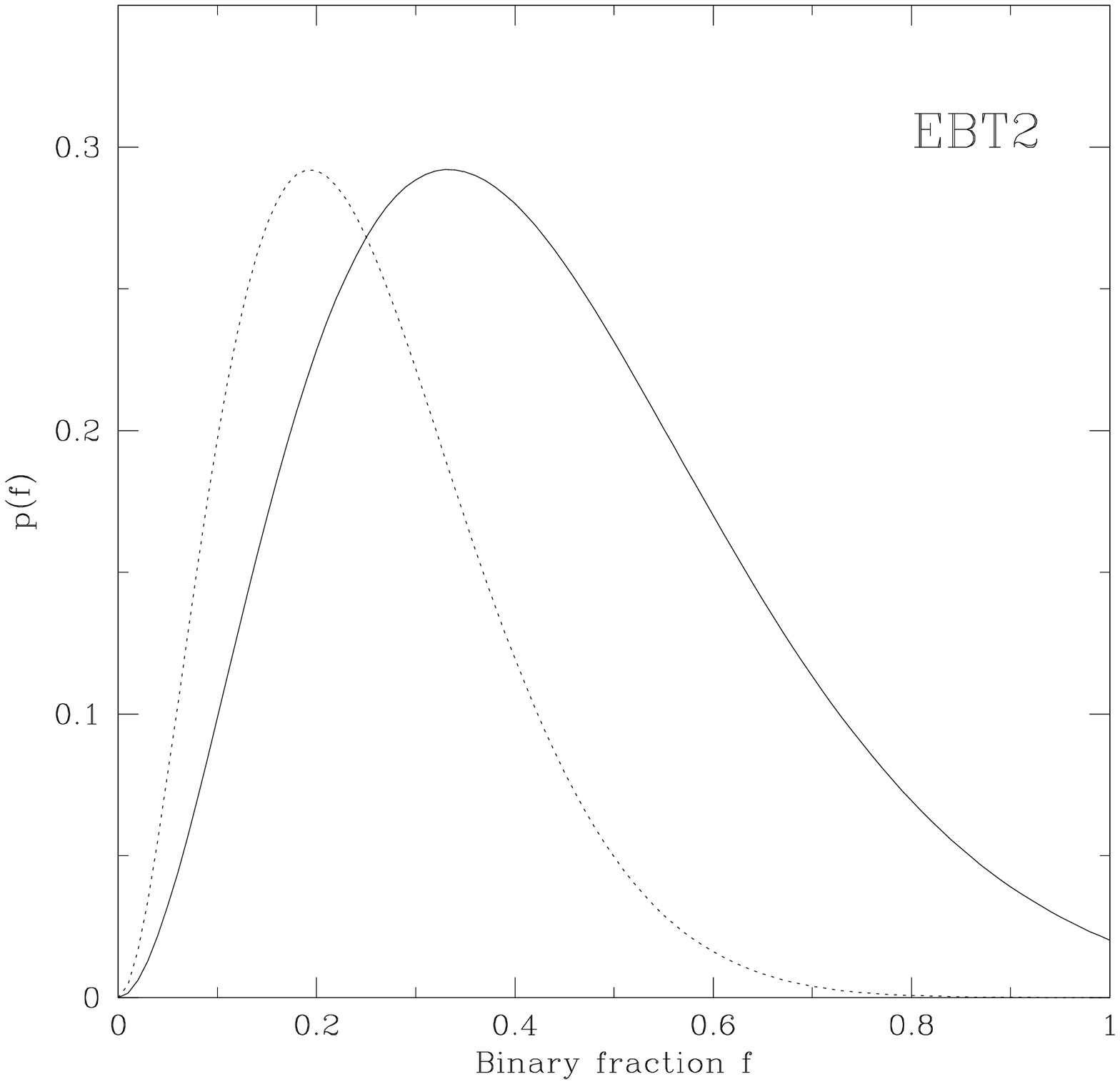}
\includegraphics[width=7.5cm]{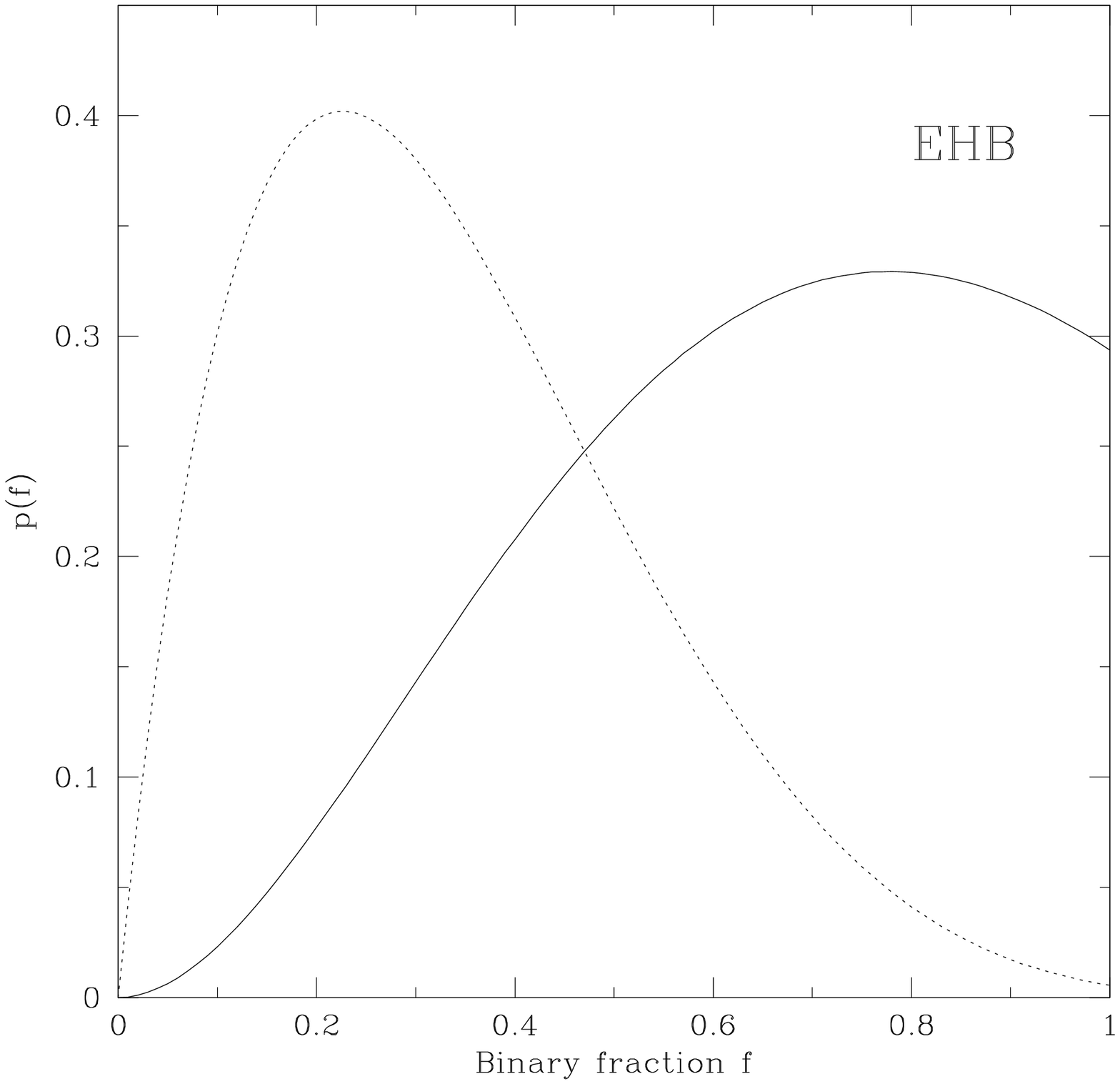}
\caption{Curve of probability for the close binary fraction $f_\mathrm{c}$ (dotted lines), and the
intermediate-period binary fraction $f_\mathrm{ip}$ (full lines). {\it Upper panel}: EBT1 stars; {\it middle panel}:
EBT2 stars; {\it Lower panel}: EHB stars.}
\label{f_frac}
\end{center}
\end{figure}

In the EBT1, we have $N$=50, and $N_B$=0 for both close and intermediate-period systems, with $\bar{d}$=0.86 and
0.48, respectively. The resulting probability curves for the two binary fractions are shown in the upper panel
of Figure~\ref{f_frac}. The curves are narrow-peaked because the efficiency of the survey is high and the sample
is sufficiently large, with the curve for $f_\mathrm{ip}$ being wider because of a lower detection probability.
As we failed to detect binaries in the EBT1, the most probable value is always zero. The probability of our
observations drops below 10\% for $f_\mathrm{c}\geq$5\% and $f_\mathrm{ip}\geq$9\%, and below 5\% for
$f_\mathrm{c}\geq$6\% and $f_\mathrm{ip}\geq$11\%. These can be adopted as upper limits at the 90\% and 95\%
confidence levels, respectively.

\subsubsection{EBT2}
\label{c_fracEBT2}

In the EBT2, we have $N$=14, and $N_B$=2 for both close and intermediate-period systems, with $\bar{d}$=0.74 and
0.43, respectively. The curves of probability for the two binary fractions are shown in the middle panel of
Figure~\ref{f_frac}. The most probable estimate of $f_\mathrm{c}$ is 19\%, and the detection probability is high
enough to obtain a relatively narrow curve: the 90\% and 95\% confidence level intervals are 6-42\% and 4-49\%,
respectively. Stronger constraints could have been obtained only with a larger sample, but this is a typical
observational limitation for GCs, because hot, isolated stars outside the inner crowded regions are few in number.
The best estimate for $f_\mathrm{ip}$ is 33\%, but the large errors affected the efficiency at detecting larger
periods, and the probability curve is very wide. The 90\% and 95\% confidence level intervals are 11-72\% and 7-85\%,
respectively. These results can serve as lower limits, since very low intermediate-period binary fractions can
be excluded.

\subsubsection{EHB}
\label{c_fracEHB}

In the EHB, we have $N$=6, $N_B$=1, and $\bar{d}$=0.74 for close binaries, while $N_B$=2 and $\bar{d}$=0.43 for
wider systems. The sample is too small to allow us to place strong constraints on the binary fractions, in
particular for $f_\mathrm{ip}$ where the detection probability is low. Therefore, we analyze this sub-group of EBT2
stars only for completeness. The more probable values are $f_\mathrm{c}$=23\% and $f_\mathrm{ip}$=77\%, and
the close binary fraction is contained within the interval 3-66\% at the 90\% confidence level, or 2-78\% at the
95\% level. Only lower limits can be obtained for $f_\mathrm{ip}$, because the probability curve does not
decrease much after its maximum: at the usual 90\% and 95\% confidence levels, $f_\mathrm{ip}\geq$24\% and
16\%, respectively.


\section{Discussion}
\label{c_discussion}

\begin{table*}
\begin{center}
\caption{Results on the close and intermediate-period binary fraction.}
\label{t_results}
\begin{tabular}{c| c c c |c c c}
\hline
\hline
 & \multicolumn{3}{|c}{Close binaries} &\multicolumn{3}{|c}{Intermediate-period binaries} \\
 & $f_\mathrm{c,max}$ & p($f_\mathrm{c})\geq$10\% & p($f_\mathrm{c})\geq$5\% &
$f_\mathrm{ip,max}$ & p($f_\mathrm{ip})\geq$10\% & p($f_\mathrm{ip})\geq$5\% \\
\hline
 & \% & \% & \% & \% & \% & \% \\
\hline
EBT1 & 0 & $\leq$5 & $\leq$6 & 0 & $\leq$9 & $\leq$11 \\
EBT2 & 19 & 6-42 & 4-49 & 33 & 11-72 & 7-85 \\
EHB  & 23 & 3-66 & 2-78 & 77 & $\geq$24 & $\geq$16 \\
\hline
\end{tabular}
\end{center}
\end{table*}

The results for the close and intermediate-period binary fraction ($f_\mathrm{c}$ and $f_\mathrm{ip}$) obtained
in Sect.\,\ref{c_fracEBT1}-\ref{c_fracEHB} are resumed in Table~\ref{t_results}, where we present the most
probable estimate
($f_\mathrm{max}$), and the 90\% and 95\% confidence level intervals (P($f$)$\geq$10\% and P($f$)$\geq$5\%)
separately for the EBT1, EBT2, and EHB stars. Despite our efforts to obtain the most precise measurements, the
very low spectral quality of the fainter stars strongly affected our results: the detection efficiency was reduced
by the large errors, especially for the longer periods, and one-fourth of the observed EBT2 stars had to be
excluded. As a consequence, the binary fractions are not very well determined, as reflected by the large
confidence intervals for the EBT2 and EHB groups. However, the statistical analysis still permits us to draw many
interesting conclusions.

The calculations of Sect.\,\ref{c_frac} are based on the fraction of detected binaries relative to the number of
stars observed in each group. Some systematic error in the estimated temperatures could thus alter the results, in
particular in the presence of stars evolving off the EHB toward higher luminosities, whose temperature could be
underestimated. The temperature and luminosity of the hotter stars are degenerate in the $V$ vs ($B-V$)
diagram, because the bluer section of the HB is almost vertical. Small luminosity variations or temperature
errors should not affect the definition of the EBT1 and EBT2 samples, because it is based on the presence of clear
photometric gaps and not on temperature boundaries. Stars ascending the asymptotic giant branch can be identified
in the $V$ vs ($U-V$) plane, where the temperature-luminosity degeneracy is much less pronounced, and EBT3 stars,
evolving directly to the WD cooling sequence, should be at all stages fainter than our magnitude limit. As a
consequence, targets assigned to the EBT2 sample should be correctly classified, while two EBT1 targets brighter
and/or bluer than the main HB population (see he upper panel of Figure~\ref{f_cmd}) could be evolved post-HB stars
originating from the
EBT2 population. One of them was excluded from the statistical analysis for other reasons (Sect.\,\ref{c_results}),
while the other one shows no sign of binarity. Assigning it to the EBT2 group, we would have $N$=15 in
Sect.\,\ref{c_fracEBT2}, causing a change of about 2\% in the values given in Table~\ref{t_results}. The results
for the EBT1 group are even less sensitive to the change, because of the larger sample. In conclusion, the
uncertainties in the definition of the EBT1 and EBT2 samples have a negligible impact on the results. In contrast,
even small systematics in the temperature scale can strongly affect the statistical analysis of the
EHB group, defined only by a temperature limit, because of the small number of observed targets. For example,
when $N$ is varied by $\pm$1, $f_\mathrm{c,max}$ and $f_\mathrm{ip,max}$ can vary by $\pm$5\% and $\pm$15\%,
respectively. These changes are not large when compared to the uncertainties given in Table~\ref{t_results}, and
the general conclusions are unaffected, but this result warns us that the results for the EHB group should be
regarded as only indicative, as already stated in Sect.\,\ref{c_fracEHB}.

Binaries with periods longer than 10~days have never been studied in GCs, and have been the target of very
few surveys even among field stars \citep[e.g.][]{MoralesRueda06}. The most probable estimate of the
intermediate-period
binary fraction among EHB stars is very high, but is most probably affected by small number statistics as can be
deduced from the very wide probability curve in the lower panel of Figure~\ref{f_frac}. However, an important result
of our investigation is a relatively high (15-20\%) lower limit for $f_\mathrm{ip}$ among EHB stars. The
estimated close binary fraction in all the clusters studied so far is either comparable to (\object{NGC\,2808},
\object{NGC\,5986}, \object{M80}) or much lower than (\object{NGC\,6752}) this value. The value of $f_\mathrm{ip}$ is
most probably higher than $f_\mathrm{c}$ even in the EBT2, because their probability curves are similar but
p($f_\mathrm{ip}$) is shifted toward higher values. The most probable values indicate that only one-fifth of the EBT2
stars can reside in close systems, but that even half of them could be binary systems with $\wp\leq$50~days. Although
we cannot provide a more reliable estimate, this result suggests that binaries wider than those investigated so far
could play an important role and should deserve more attention, particularly where close systems are lacking.

A second prominent result clearly evident from Table~\ref{t_results} is that the 90\% confidence intervals in the EBT1
and EBT2 do not overlap, for both $f_\mathrm{c}$ and $f_\mathrm{ip}$. This means that the probability that the
binary fraction has the same value in these two sections of the HB is negligible (of the order of 1\%): {\it the G1
gap thus appears as a discontinuity in the binary fraction}, there being a very small quantity of binaries among the
stars cooler than the gap, and about 20-30\% for hotter objects. Unfortunately, our data cannot exclude $f_\mathrm{c}$
and $f_\mathrm{ip}$ monotonically increasing with the temperature rather than showing a real
discontinuity at G1. In this regard, we note that there is reasonable agreement between the results in the
EBT2 and in its (hotter)
sub-group the EHB, in particular for the close binary fraction (compare also the shapes of the
dotted curves in the middle and lower panels of Figure~\ref{f_frac}). Thus, $f_\mathrm{c}$ could be quite homogeneous
within the EBT2 population, instead of increasing with temperature, but no firm conclusion can be drawn because of the
too wide confidence intervals. Future observations should help us to clarify this issue: a sudden increase in the
binarity in correspondence with G1 would strongly relate it to the formation of all EBT2 stars, while a $f_\mathrm{c}$
slowly increasing with temperature may indicate that the progeny of close systems are preferentially hotter.

In the context of the He-enrichment scenario for the formation of hot HB stars in GCs, the difference in binary
fraction among EBT1 and EBT2 stars contrasts with the investigation of \citet{DOrazi10} in \object{M4}, who found a
much lower quantity of binaries among red giants that displayed evidence of chemical enrichment, than among normal
ones. They argue that this difference is naturally explained by assuming that the second stellar
generation formed in a denser environment, where more frequent dynamical interactions
enhanced the disruption rate of binary systems. However, in the multi-population models of \object{NGC\,2808}, the
EBT2 is interpreted as the progeny of the latest and He-richest of the three populations observed in the MS
\citep{DAntona05,DAlessandro2010} and, following \citet{DOrazi10}, we would expect the EBT2 to be depleted in binaries,
at variance with what is observed. If their results were confirmed as a general behavior of the chemically polluted
stars in GCs, our results would argue against the link between the EBT2 and the He-enriched stars. However, an
alternative interpretation would be that both the He-enrichment and the binary scenarios co-exist in the cluster,
as different channels for the formation of blue HB stars. In this case, in the EBT2 both the progeny of He-enriched
stars and products of binary interactions would be found. This would cause a higher frequency of EBT2 stars in the HB
with respect to the fraction of He-rich MS stars, but binaries in GCs usually represent a minor fraction of the
entire population, and this difference could pass unnoticed.
No splitting of the red- or sub- giant branch has been detected so far in \object{NGC\,2808}, but this
cluster shows a strong Na-O anticorrelation \citep{Carretta06}, which has often been interpreted as a consequence
of cluster self-pollution. It would therefore be really instructive to repeat the D'Orazi et al.'s investigation
among the three different groups of red giants identified by \citet{Carretta06} on the basis of their Na/O value,
for a direct comparison with the values presented in Table~\ref{t_results}.

Previous surveys focused mainly on the close binary fraction among EHB stars, defined by the temperature boundary
of 20\,000~K. This is more than a conventional definition because, as noted in Sect.\,\ref{c_intro}, the post-HB
evolution of EHB stars is very distinct from that of cooler stars. However, the formation mechanism does not
necessarily differ, and population synthesis models show that sdB stars as cool as 15\,000-16\,000~K can be
formed with the same binary channels used to model the field EHB star population \citep{Han03}. In addition,
massive (0.75~M$_\odot$) EHB stars move to cooler temperatures in the first stages of the post-HB evolution
\citep{Han02}. The previous investigations were not limited to stars hotter than 20\,000~K, and the only
close system found in \object{M80}, which \citet{Moni09} included in a statistical analysis similar to ours, is
actually cooler than this temperature ($\sim$18\,000~K). For all these reasons, we can safely use our estimate of
the close binary fraction in the EBT2 population for comparison with both the results for other clusters, and
the expectations of the binary model for sdB star formation. We also recall that the adopted canonical
temperature scale might not be appropriate, as the EBT2 and EHB coincide in the model of \citet{DAlessandro2010}.

The most reliable estimate of $f_\mathrm{c}$ is low compared to field values, as observed in all the GCs observed
so far. Even the lowest fractions proposed in the literature for the field sdB population are very unlikely
\citep[$\sim$45\%, e.g.][]{Napiwotzki04,MoralesRueda06}, and the highest ones can safely be excluded
\citep[$\sim$70\%,][]{Maxted01}. However, even the extremely low $f_\mathrm{c}$ measured in \object{NGC\,6752}
\citep[4\%,][]{Moni08a} can be excluded with a 95\% confidence level. We thus find that \object{NGC\,2808} is
similar to \object{NGC\,5986}, where \citet{Moni09} estimated that $f_\mathrm{c}$=25\%, although with great
uncertainties. The similarity can be drawn even further, because both these clusters are 2-3~Gyr younger than
\object{NGC\,6752} \citep{DeAngeli05}. While the results of \citet{Moni09} had to be confirmed, the probability
that the close binary fraction is equal or very similar in all these three clusters is negligible, and the trend
of $f_\mathrm{c}$ with age depicted in Figure~1 of \citet{Moni10} is likely real, following the close binary
fraction-age relation proposed by \citet{Moni08a} and modeled by \citet{Han08} in the context of the binary
scenario. On the other hand, \object{NGC\,2808} also confirms the problems of this scenario pointed out by
\citet{Moni09}: no model proposed by \citet{Han08} can predict both the very low $f_\mathrm{c}$ of
\object{NGC\,6752} and a steep increase with decreasing age up to $f_\mathrm{c}\approx$20\% for clusters 2-3 Gyr younger.
As can be seen from their Figure~3, the proposed solutions do not have a sufficiently steep gradient, and all
the curves with $f_\mathrm{c}$(13 Gyr)$\leq$5\% are nearly flat for old populations, so that
$f_\mathrm{c}\leq$10\% for any cluster older than 8 Gyr. Nevertheless, the simulations of
\citet{Han08} depend strongly on a set of poorly known parameters. In particular, \citet{Marsh95} deduced a
very high CE ejection efficiency ($\alpha_\mathrm{CE}\approx$1) from their observations of double-degenerate
close binaries, but a much lower value (0.2-0.3) was proposed by \citet{Zorotovic10} as a result of their study
of a large sample of post-CE binaries. \citet{Han08} assumed a $\alpha_\mathrm{CE}$=0.5-1.0, but from their
Figure~3 it can be seen that lower values tend to return a steeper slope of the $f_\mathrm{c}$-age relation,
although still predicting too high values at 13 Gyr. It is therefore possible that, with a different set
of parameters, the binary scenario can be reconciled with the observations.


\section{Conclusions}
\label{c_conclusions}

We have analyzed the radial velocity variations of a sample of 50 EBT1 stars (T$_\mathrm{eff}$=12\,000-17\,000~K)
and 14 EBT2 (T$_\mathrm{eff}$=17\,000-22\,000~K) stars in the horizontal branch of \object{NGC\,2808}, which
were spectroscopically observed in ten epochs spanning a temporal interval of $\sim$75~days. We detected two
close binaries (period $\wp\leq$10~days), and two wider systems ($\wp\geq$10~days and $\wp$=5-20~days, respectively).
All these object belong to the EBT2 population, while no binary was detected among the cooler EBT1 stars.
We estimated the most probable fraction of close ($\wp\leq$10~days) and intermediate-period ($\wp$=10-200~days)
binaries, $f_\mathrm{c}$ and $f_\mathrm{ip}$, plus the ranges corresponding to a 90\% and 95\% confidence
level, respectively, in both the EBT1 and EBT2 groups of stars. Although for hot stars both the sample and the
survey efficiency were reduced by the low S/N of the spectra, we were able to draw some important conclusions:
\begin{itemize}
\item We found a relatively high lower limit (15-20\%) for $f_\mathrm{ip}$ among EHB stars
(T$_\mathrm{eff}\geq$20\,000~K). The intermediate-period binary fraction is most probably higher than
$f_\mathrm{c}$ even for EBT2 stars, where only one-fifth of stars could reside in close systems, but up to half
of them could be binaries of period shorter than 50 days. We cannot obtain a better estimate, but this result
warns that these so-far unstudied systems could play an important role, in particular in clusters with a strong
lack of close EHB binaries (e.g. \object{NGC\,6752}).

\item The G1 gap separating the EBT1 and the EBT2 in the CMD appears as a discontinuity for both the close and the
intermediate-period binary fraction, because the probability that either $f_\mathrm{c}$ or $f_\mathrm{ip}$
is the same in the two populations is negligible. Nevertheless, a smooth increase in binarity with temperature,
rather than a discontinuity, cannot be excluded by our observations. If the observed higher binary frequency
among chemically unpolluted RGB stars in \object{M4} \citep{DOrazi10} should be confirmed as a general behavior of
multi-populations in GCs, this could represent a problem for the frequently proposed scenario in which
the EBT2 stars are the progeny of the most He-enriched stellar generation observed in the MS.

\item A value of $f_\mathrm{c}$ as high as that of field stars can be excluded, but close EHB systems are surely
much more frequent than in \object{NGC\,6752}. In contrast, for \object{NGC\,2808} we find results very similar
to \object{NGC5986}. Both these clusters are 2-3 Gyr younger than \object{NGC\,6752}, which indicates that an
$f_\mathrm{c}$-age is present in GCs.

\item The similar $f_\mathrm{c}$ found in \object{NGC\,2808} and \object{NGC\,5986} also confirms the problems of
reconciling the binary scenario with the observations in GCs, because the predictions proposed so far cannot
account for both the extremely low $f_\mathrm{c}$ of \object{NGC\,6752}, and the much higher values found in
younger clusters. We argue, however, that a change in the model parameters could enable closer agreement to be
reached.
\end{itemize}


\begin{acknowledgements}
CMB and SV acknowledge the Chilean Centro de Excelencia en Astrof\'isica y Tecnolog\'ias Afines (CATA). 
\end{acknowledgements}


\bibliographystyle{aa}
\bibliography{biblio2808}

\begin{table*}
\begin{center}
\caption{Data of program stars.}
\label{t_datatarg}
\begin{tabular}{c c c c c c c c}
\hline
\hline
ID & RA & dec & $V$ & ($U-V$) & T$_\mathrm{eff}$ & $\overline{\mathrm RV}$ & $\Delta_\mathrm{max}$ \\
 & hh:mm:ss & $^{\circ}$: ' : '' & & & K &~km~s$^{-1}$ &~km~s$^{-1}$ \\
\hline
15924 & 9:11:41.89 & $-$64:42:05.7 & 17.975 & $-$0.932 & 14600 & - & - \\
54434 & 9:12:07.09 & $-$64:50:31.4 & 17.890 & $-$0.371 & 14300 & 103.0$\pm$0.4 & 1.18 \\
9715  & 9:11:59.57 & $-$64:47:44.3 & 18.772 & $-$0.497 & 18100 & 97.2$\pm$1.6 & 5.15 \\
53800 & 9:11:57.47 & $-$64:50:41.2 & 17.604 & $-$0.352 & 13200 & 94.0$\pm$0.5 & 2.76 \\
14923 & 9:11:27.83 & $-$64:46:22.5 & 17.471 & $-$0.260 & 12700 & 85.7$\pm$1.0 & 2.24 \\
7700  & 9:11:52.41 & $-$64:49:18.1 & 19.168 & $-$0.510 & 20500 & 108.3$\pm$2.4 & 5.80 \\
13405 & 9:11:41.82 & $-$64:49:16.1 & 17.807 & $-$0.289 & 14000 & 95.7$\pm$1.0 & 2.53 \\
7227  & 9:12:06.11 & $-$64:49:32.6 & 18.172 & $-$0.403 & 15400 & 94.1$\pm$0.8 & 1.75 \\
8607  & 9:12:12.06 & $-$64:48:42.8 & 17.545 & $-$0.254 & 13000 & 101.9$\pm$0.7 & 2.02 \\
57252 & 9:12:09.32 & $-$64:49:43.1 & 17.450 & $-$0.219 & 12700 & 93.7$\pm$0.6 & 1.61 \\
53504 & 9:12:03.67 & $-$64:50:45.8 & 18.348 & $-$0.478 & 16100 & 95.5$\pm$1.1 & 2.43 \\
10078 & 9:12:17.15 & $-$64:47:17.1 & 18.894 & $-$0.573 & 18800 & 88.5$\pm$2.1 & 1.80 \\
11623 & 9:12:59.52 & $-$64:42:51.5 & 18.997 & $-$0.659 & 19400 & - & - \\
9121  & 9:12:21.71 & $-$64:48:18.3 & 18.296 & $-$0.491 & 15900 & 91.7$\pm$1.2 & 2.30 \\
10620 & 9:12:29.08 & $-$64:46:20.8 & 17.956 & $-$0.369 & 14500 & 92.0$\pm$0.9 & 2.87 \\
10719 & 9:12:24.79 & $-$64:46:09.9 & 17.795 & $-$0.330 & 13900 & 101.2$\pm$1.1 & 1.48 \\
51625 & 9:12:08.02 & $-$64:51:15.8 & 17.750 & $-$0.439 & 13800 & 100.2$\pm$0.5 & 2.84 \\
51455 & 9:11:58.74 & $-$64:51:18.5 & 17.422 & $-$0.294 & 12600 & 90.1$\pm$1.4 & 1.96 \\
49247 & 9:12:08.85 & $-$64:51:53.0 & 17.519 & $-$0.243 & 12900 & - & - \\
50965 & 9:12:24.44 & $-$64:51:26.1 & 17.714 & $-$0.288 & 13600 & 81.3$\pm$1.4 & 1.96 \\
47936 & 9:12:07.89 & $-$64:52:13.3 & 17.902 & $-$0.465 & 14400 & 93.3$\pm$0.5 & 1.57 \\
47593 & 9:12:09.69 & $-$64:52:18.3 & 17.632 & $-$0.312 & 13300 & 93.0$\pm$1.0 & 0.91 \\
53681 & 9:12:22.70 & $-$64:50:42.8 & 19.095 & $-$0.620 & 20100 & 89.8$\pm$1.8 & 1.82 \\
53679 & 9:12:11.84 & $-$64:50:43.0 & 17.470 & $-$0.178 & 12700 & 92.1$\pm$0.7 & 1.71 \\
55759 & 9:12:32.23 & $-$64:50:08.8 & 19.064 & $-$0.628 & 20000 & 79.2$\pm$2.8 & 4.04 \\
52126 & 9:12:11.18 & $-$64:51:07.6 & 18.098 & $-$0.386 & 15100 & 89.4$\pm$0.9 & 2.00 \\
55692 & 9:12:15.38 & $-$64:50:10.2 & 17.804 & $-$0.297 & 13900 & 98.4$\pm$0.6 & 2.55 \\
9519  & 9:12:29.66 & $-$64:47:56.3 & 19.271 & $-$0.619 & 21200 & 99.2$\pm$1.9 & 7.95 \\
47443 & 9:12:33.43 & $-$64:52:20.4 & 17.751 & $-$0.374 & 13800 & 100.1$\pm$0.9 & 2.46 \\
48326 & 9:12:23.85 & $-$64:52:07.1 & 18.846 & $-$0.673 & 18600 & 95.4$\pm$1.4 & 2.68 \\
50088 & 9:12:37.34 & $-$64:51:39.8 & 17.795 & $-$0.348 & 13900 & 95.9$\pm$0.8 & 2.02 \\
50410 & 9:12:28.99 & $-$64:51:35.0 & 18.232 & $-$0.453 & 15600 & - & - \\
52303 & 9:12:53.68 & $-$64:51:03.8 & 18.872 & $-$0.532 & 18700 & 94.5$\pm$1.5 & 2.81 \\
51077 & 9:12:14.46 & $-$64:51:24.4 & 18.978 & $-$0.740 & 19300 & 102.8$\pm$1.3 & 1.46 \\
46155 & 9:12:18.44 & $-$64:52:42.3 & 18.169 & $-$0.506 & 15400 & 94.7$\pm$1.0 & 2.13 \\
44537 & 9:12:20.59 & $-$64:53:09.6 & 19.249 & $-$0.661 & 21100 & 102.8$\pm$2.2 & 1.88 \\
41203 & 9:12:31.09 & $-$64:54:14.5 & 18.063 & $-$0.398 & 14900 & 99.6$\pm$1.0 & 1.40 \\
46960 & 9:12:29.17 & $-$64:52:28.7 & 17.840 & $-$0.364 & 14100 & 94.7$\pm$1.2 & 1.13 \\
46398 & 9:12:22.04 & $-$64:52:38.1 & 17.585 & $-$0.695 & 13200 & 86.3$\pm$1.5 & 2.03 \\
50078 & 9:12:20.46 & $-$64:51:40.3 & 18.014 & $-$0.368 & 14700 & - & - \\
43422 & 9:12:06.35 & $-$64:53:29.6 & 18.028 & $-$0.452 & 14800 & 98.1$\pm$1.9 & 0.97 \\
45166 & 9:12:17.45 & $-$64:52:59.0 & 17.626 & $-$0.422 & 13300 & 82.8$\pm$1.5 & 2.26 \\
46460 & 9:12:11.77 & $-$64:52:37.3 & 18.029 & $-$0.349 & 14800 & 98.5$\pm$0.7 & 1.54 \\
45677 & 9:12:02.04 & $-$64:52:50.4 & 17.582 & $-$0.386 & 13200 & 99.4$\pm$0.7 & 1.12 \\
46512 & 9:12:08.63 & $-$64:52:36.6 & 17.690 & $-$0.373 & 13500 & 92.9$\pm$0.8 & 0.95 \\
37288 & 9:12:34.35 & $-$64:57:33.5 & 17.691 & $-$0.394 & 13500 & 92.1$\pm$0.8 & 2.65 \\
40263 & 9:12:07.13 & $-$64:54:41.3 & 19.395 & $-$0.735 & 22100 & - & - \\
37744 & 9:11:58.76 & $-$64:56:45.5 & 19.229 & $-$0.542 & 21100 & 100.2$\pm$1.7 & 1.88 \\
37345 & 9:12:12.46 & $-$64:57:27.0 & 17.916 &    0.068 & 14400 & $\sim$150 & - \\
39797 & 9:12:02.88 & $-$64:54:55.8 & 17.778 & $-$0.381 & 13900 & 87.6$\pm$0.8 & 1.57 \\
39922 & 9:12:10.73 & $-$64:54:51.2 & 17.412 & $-$0.228 & 12500 & 86.2$\pm$0.5 & 1.67 \\
58322 & 9:12:34.08 & $-$65:01:05.8 & 17.683 & $-$0.413 & 13500 & 92.3$\pm$0.7 & 2.29 \\
47336 & 9:12:04.59 & $-$64:52:22.7 & 17.782 & $-$0.237 & 13900 & 88.5$\pm$0.8 & 0.87 \\
42482 & 9:12:05.53 & $-$64:53:47.3 & 18.712 & $-$0.893 & 17900 & 93.6$\pm$1.5 & 1.60 \\
44070 & 9:12:00.17 & $-$64:53:18.0 & 18.567 & $-$0.395 & 17100 & 98.0$\pm$1.1 & 1.79 \\
45042 & 9:11:58.06 & $-$64:53:01.3 & 17.969 & $-$0.418 & 14600 & 89.5$\pm$0.8 & 2.73 \\
43840 & 9:12:03.54 & $-$64:53:21.9 & 17.698 & $-$0.386 & 13600 & 97.3$\pm$0.5 & 2.05 \\
44295 & 9:11:55.54 & $-$64:53:13.9 & 17.507 & $-$0.326 & 12900 & 93.4$\pm$0.5 & 2.61 \\
39744 & 9:11:46.36 & $-$64:54:57.9 & 17.402 & $-$0.288 & 12500 & 92.7$\pm$0.8 & 2.25 \\
39433 & 9:12:01.85 & $-$64:55:09.6 & 17.915 & $-$0.477 & 14400 & - & - \\
\hline
\end{tabular}
\end{center}
\end{table*}

\begin{table*}
\begin{center}
\caption{Data of program stars.}
\label{t_datatarg}
\begin{tabular}{c c c c c c c c}
\hline
\hline
ID & RA & dec & $V$ & ($U-V$) & T$_\mathrm{eff}$ & $\overline{\mathrm RV}$ & $\Delta_\mathrm{max}$ \\
 & hh:mm:ss & $^{\circ}$: ' : '' & & & K &~km~s$^{-1}$ &~km~s$^{-1}$ \\
\hline
43287 & 9:11:53.32 & $-$64:53:32.1 & 18.913 & $-$0.464 & 18900 & 86.1$\pm$1.4 & 0.60 \\
49443 & 9:11:51.16 & $-$64:51:50.0 & 17.460 & $-$0.266 & 12700 & 87.1$\pm$1.1 & 1.60 \\
47972 & 9:11:54.83 & $-$64:52:12.8 & 17.514 & $-$0.262 & 12900 & 97.5$\pm$0.6 & 2.34 \\
30033 & 9:11:33.68 & $-$64:56:24.9 & 19.269 & $-$0.794 & 21200 & - & - \\
43170 & 9:11:48.45 & $-$64:53:34.1 & 17.649 & $-$0.333 & 13400 & 88.0$\pm$0.5 & 1.92 \\
47037 & 9:11:54.86 & $-$64:52:27.7 & 17.389 & $-$0.263 & 12400 & 90.1$\pm$0.6 & 2.27 \\
33067 & 9:11:42.45 & $-$64:52:23.4 & 17.740 & $-$0.325 & 13700 & 96.5$\pm$0.8 & 1.98 \\
32516 & 9:11:16.93 & $-$64:52:53.6 & 17.508 & $-$0.262 & 12900 & 92.6$\pm$1.0 & 2.08 \\
32670 & 9:11:24.84 & $-$64:52:45.5 & 18.509 & $-$0.730 & 16800 & 88.2$\pm$2.0 & 4.74 \\
45980 & 9:11:56.65 & $-$64:52:45.4 & 18.354 & $-$0.363 & 16100 & 90.1$\pm$0.9 & 2.28 \\
49321 & 9:11:54.05 & $-$64:51:52.0 & 18.029 & $-$0.498 & 14800 & 85.5$\pm$0.6 & 2.37 \\
41115 & 9:11:52.59 & $-$64:54:17.0 & 17.782 & $-$0.435 & 13900 & 90.8$\pm$0.7 & 2.03 \\
50769 & 9:11:44.40 & $-$64:51:29.6 & 17.859 & $-$0.383 & 14200 & 89.6$\pm$0.7 & 1.75 \\
35362 & 9:11:19.47 & $-$64:50:04.2 & 17.882 & $-$0.243 & 14200 & 92.9$\pm$1.1 & 1.59 \\
34446 & 9:11:31.00 & $-$64:51:04.4 & 18.962 & $-$0.539 & 19200 & 91.4$\pm$2.1 & 1.82 \\
32888 & 9:11:18.78 & $-$64:52:32.7 & 18.315 & $-$0.370 & 15900 & - & - \\
54539 & 9:11:50.81 & $-$64:50:29.9 & 18.906 & $-$0.885 & 18900 & - & - \\
54675 & 9:11:46.80 & $-$64:50:27.4 & 17.516 & $-$0.345 & 12900 & 84.3$\pm$1.0 & 1.86 \\
55009 & 9:11:52.90 & $-$64:50:22.1 & 17.807 & $-$0.325 & 14000 & 86.0$\pm$0.9 & 0.83 \\
13467 & 9:11:22.19 & $-$64:49:11.5 & 18.996 & $-$0.558 & 19400 & - & - \\
53751 & 9:11:45.60 & $-$64:50:41.9 & 17.597 & $-$0.268 & 13200 & 88.2$\pm$1.2 & 1.18 \\
51656 & 9:11:48.40 & $-$64:51:15.3 & 17.754 & $-$0.349 & 13800 & 91.0$\pm$0.7 & 1.55 \\
52012 & 9:11:51.63 & $-$64:51:09.6 & 17.414 & $-$0.295 & 12600 & 86.8$\pm$0.8 & 1.06 \\
\hline
\end{tabular}
\end{center}
\end{table*}

\end{document}